\documentclass[aps, reprint, prd, superscriptaddress, nofootinbib, notitlepage, amsmath, floatfix, longbibliography,preprintnumbers]{revtex4-2}
\usepackage{xcolor}
\usepackage{graphicx}
\usepackage{multirow}
\usepackage[colorlinks=true, linkcolor=blue, citecolor=blue, urlcolor=blue]{hyperref}

\newcommand{\alphas}{\alpha_{{\rm s}}}

\newcommand{\zs}{z_{{\rm s}}}
\newcommand{\ts}{t_{{\rm s}}}

\begin{document}
\raggedbottom

\preprint{FERMILAB-PUB-26-0796-T}

\title{Unpolarized proton PDF with physical quark masses and large momenta}


\author{Xiang Gao}
\affiliation{Physics Division, Brookhaven National Laboratory, Upton, New York 11973, USA}

\author{Andrew Hanlon}
\affiliation{Center for Nuclear Research, Department of Physics, Kent State University, Kent, Ohio 44242, USA}

\author{Jinchen He}
\affiliation{Fermi National Accelerator Laboratory, Batavia, Illinois 60510, USA}
\affiliation{Maryland Center for Fundamental Physics, University of Maryland, College Park, MD 20742, USA}
\affiliation{Physics Division, Argonne National Laboratory, Lemont, Illinois 60439, USA}

\author{Xiangyu Jiang}
\affiliation{Physics Department, Indiana University, Bloomington, Indiana 47405, USA}

\author{Swagato Mukherjee}
\affiliation{Physics Division, Brookhaven National Laboratory, Upton, New York 11973, USA}

\author{Peter Petreczky}
\affiliation{Physics Division, Brookhaven National Laboratory, Upton, New York 11973, USA}

\author{Qi Shi}
\email{qshi3@kent.edu}
\affiliation{Center for Nuclear Research, Department of Physics, Kent State University, Kent, Ohio 44242, USA}

\author{Yong Zhao}
\affiliation{Physics Division, Argonne National Laboratory, Lemont, Illinois 60439, USA}

\begin{abstract}
We present a lattice quantum chromodynamics (QCD) calculation of the proton isovector ($u-d$) unpolarized parton distribution function (PDF) at physical sea and valence quark masses, using Coulomb-gauge-fixed quark bilinears within the large-momentum effective theory framework. The calculation is performed on a (2+1)-flavor ensemble with a lattice spacing of $a=0.076$~fm and proton momenta up to $2.3$~GeV. We combine kinematically enhanced nucleon interpolating operators, the Coulomb-gauge method, and high-statistics measurements to improve the signal quality at large momenta and quark separations. We use the static regularization-independent momentum subtraction and hybrid schemes to renormalize the bare matrix elements. 
The light-cone PDF in the $\overline{\rm MS}$ scheme is obtained from the quasi-PDF through next-to-leading-order (NLO) perturbative matching supplemented by next-to-leading-logarithmic renormalization-group resummation. With reduced uncertainties, the resulting PDF is consistent with our previous calculation on a similar lattice ensemble with $a=0.06$~fm and a valence pion mass of 300 MeV. Notably, despite the reduced uncertainties, the agreement with the phenomenological PDF persists over the longitudinal momentum fraction range $0.25\lesssim|x|\lesssim0.75$, where the lattice calculation is expected to be reliable.
This agreement is further supported by a comparison in coordinate space, where our short-distance lattice matrix elements are consistent with those obtained from the phenomenological PDF through NLO matching. In particular, we observe close agreement in the imaginary part for the first time. Together, these results represent an encouraging step toward precision benchmarking of lattice QCD calculations of PDFs.

\end{abstract}
 
\maketitle

\section{\label{sec:intro}Introduction}
Understanding the internal structure of the nucleon remains one of the defining challenges of quantum chromodynamics (QCD). The longitudinal momentum distributions of quarks and gluons are encoded in parton distribution functions (PDFs), which serve as indispensable nonperturbative inputs for predictions of high-energy processes. Phenomenological PDFs are primarily determined through global QCD analyses of deep-inelastic scattering (DIS) and other hard-scattering data~\cite{Alekhin:2017kpj,Hou:2019efy,Bailey:2020ooq,NNPDF:2021njg}. A first-principles theoretical determination can complement these analyses. With the ongoing 12~GeV program at the Thomas Jefferson National Accelerator Facility (JLab)~\cite{Dudek:2012vr} and the forthcoming Electron-Ion Collider (EIC)~\cite{Accardi:2012qut,AbdulKhalek:2021gbh}, experimental explorations of nucleon structure will reach unprecedented precision, demanding commensurate advancements in first-principles theoretical calculations~\cite{Amoroso:2022eow}.

Lattice QCD provides a first-principles, nonperturbative approach to this task. Traditional lattice calculations, however, were long restricted to the lowest Mellin moments of the PDFs~\cite{Hagler:2009ni}. Large-momentum effective theory (LaMET) provides a way to access the $x$ dependence~\cite{Ji:2013dva,Ji:2014gla,Ji:2020ect} through quasi-PDFs defined from equal-time spatial correlations in a boosted hadron. The light-cone PDFs are obtained from the quasi-PDFs through a large-momentum expansion and perturbative matching~\cite{Ji:2022ezo,Ji:2024oka}. Complementary Euclidean approaches include hadronic tensor~\cite{Liu:1993cv}, Compton amplitude~\cite{Chambers:2017dov}, current-current correlators~\cite{Braun:2007wv,Ma:2017pxb}, pseudo-PDF~\cite{Radyushkin:2017cyf}, the heavy-quark operator product expansion~\cite{Detmold:2005gg,Detmold:2021uru}, and methods for higher Mellin moments~\cite{Davoudi:2012ya,Shindler:2023xpd}. Building on significant theoretical developments over the past decade, LaMET has transformed the direct lattice calculation of $x$-dependent partonic distributions into an active precision frontier of QCD studies~\cite{Constantinou:2020hdm,Zhao:2025oto}, with a broad body of nucleon results across the unpolarized, helicity, and transversity PDFs~\cite{Chen:2016utp, Lin:2017ani, Chen:2018xof, Alexandrou:2018pbm, LatticeParton:2018gjr, Alexandrou:2018eet, Lin:2018pvv, Alexandrou:2019lfo, Fan:2020nzz, Alexandrou:2020zbe, Alexandrou:2020qtt, Lin:2020fsj, Gao:2022uhg, Alexandrou:2021oih, Gao:2023ktu, Holligan:2023jqh, Holligan:2024wpv, Gao:2026wlz, Gao:2026hix}. As such, benchmarking PDF calculations, particularly those within the LaMET framework, is becoming increasingly important for establishing the broader and long-term impact of lattice QCD as a first-principles tool for studying parton physics, as emphasized in Refs.~\cite{Lin:2017snn,Constantinou:2020hdm}.

The proton isovector unpolarized PDF, $f_u(x)-f_d(x)$, provides the ideal channel for this purpose. Thanks to decades of sustained efforts and the accumulation of high-precision experimental data, the global analyses of proton unpolarized quark PDFs have achieved remarkable precision in the valence region, with close agreement among independent groups~\cite{Hou:2019efy,Bailey:2020ooq,NNPDF:2021njg}. Meanwhile, the isovector combination $(u-d)$ provides a particularly clean benchmark for lattice QCD calculations of both the Mellin moments and $x$-dependence of PDFs~\cite{Lin:2017snn,Constantinou:2020hdm}, thanks to the cancellation of quark-gluon mixing and disconnected quark-loop contributions.

While there have been quite a few lattice calculations of the proton isovector unpolarized PDF~\cite{Lin:2014zya,Alexandrou:2015rja,Chen:2016utp,Alexandrou:2016jqi,Chen:2017mzz,Lin:2017ani,Orginos:2017kos,Alexandrou:2018pbm,Chen:2018xof,LatticeParton:2018gjr,Alexandrou:2019lfo,Joo:2019jct,Bhat:2020ktg,Joo:2020spy,Alexandrou:2020qtt,Lin:2020fsj,Fan:2020nzz,Egerer:2021ymv,Karpie:2021pap,Bhat:2022zrw,Gao:2022uhg}, most either suffer from substantial uncertainties or fail to agree with global analyses within $1\sigma$ in the moderate-$x$ region. Such discrepancies have been observed in calculations based on both LaMET~\footnote{It is worth noting that most LaMET calculations were carried out before the development of modern analysis techniques such as hybrid-scheme renormalization~\cite{Ji:2020brr,LatticePartonLPC:2021gpi}, leading-renormalon resummation~\cite{Zhang:2023bxs,Holligan:2023rex}, renormalization group~\cite{Gao:2021hxl,Su:2022fiu} and threshold resummations~\cite{Ji:2023pba,Ji:2024hit}, and asymptotic extrapolation~\cite{Ji:2026vir}. Although these techniques play important roles in reducing theoretical uncertainties, they alone are insufficient to resolve some of the discrepancies observed in the resulting PDFs.} and short-distance expansion approaches. By separating the PDF into the valence $(q-\bar{q})$ and full-quark $(q+\bar{q})$ channels, corresponding to the real and imaginary parts of the equal-time correlator, respectively, some previous studies have found agreement in the real part between lattice and global analyses~\cite{Alexandrou:2020qtt,Fan:2020nzz,Gao:2022uhg}, whereas the imaginary part consistently exhibits deviations~\cite{Chen:2017mzz,Alexandrou:2020qtt,Fan:2020nzz,Gao:2022uhg}. At short distances, the operator product expansion (OPE)~\cite{Izubuchi:2018srq} reveals that the leading nonperturbative contribution to the imaginary part is governed by the momentum fraction $\langle x\rangle$. Therefore, the observed deviation is closely related to the discrepancy in the isovector $\langle x\rangle$, which has also been observed in direct lattice calculations of Mellin moments~\cite{Bali:2018zgl,Alexandrou:2019ali,Harris:2019bih,Alexandrou:2020sml,Mondal:2020cmt}. 
Dedicated efforts to control the lattice systematics have brought the isovector $\langle x\rangle$ into agreement with phenomenology~\cite{Mondal:2020ela,Mondal:2021oot,Djukanovic:2024krw,Alexandrou:2026oks}, but comparable efforts have yet to be undertaken for calculations of the PDF. Nonetheless, Ref.~\cite{Ji:2026pen} reanalyzed the proton unpolarized quark PDF data from four independent lattice groups~\cite{LatticeParton:2018gjr,Alexandrou:2020qtt,Lin:2020fsj,Chen:2024rgi}. By fixing the isovector $\langle x\rangle$ in the OPE of the imaginary matrix elements to a lattice determination~\cite{Djukanovic:2024krw} consistent with phenomenology, as well as incorporating state-of-the-art LaMET analysis techniques, the study found overall good agreement with global analyses in the moderate-$x$ region across all four lattice datasets. This work demonstrates the potential for successful benchmarking of LaMET determinations of PDFs against phenomenology once the remaining systematic uncertainties in lattice matrix elements are brought under control.

Recently, Coulomb-gauge (CG) correlators have been proposed as an alternative to gauge-invariant (GI) Wilson-line operators for calculating parton distributions within the LaMET framework~\cite{Gao:2023lny,Zhao:2023ptv}. The quasi-PDFs defined from these correlators belong to the same universality class of LaMET~\cite{Hatta:2013gta,Ji:2020ect}, which can be perturbatively matched onto the same light-cone PDFs. Thanks to the absence of Wilson lines and the associated linear power divergences~\cite{Gao:2023lny,Zhang:2024omt}, the renormalization of CG correlators becomes much simpler, which also leads to enhanced precision in the long-distance matrix elements. Besides, the preservation of 3D rotational symmetry allows for reaching larger off-axis momenta without increasing the statistical errors. Due to these nice features, we explored the CG method for the first time in a calculation of the proton isovector PDFs on a fine lattice with spacing $a=0.06$~fm and valence pion mass of $300$ MeV~\cite{Gao:2026hix}, using up-to-date LaMET analysis techniques. Working at two large off-axis proton momenta, $P_z=2.43$ and $3.04$~GeV, this study obtained the isovector unpolarized PDF compatible with phenomenological results at moderate $x$ within $1\sigma$. However, the uncertainties are still large in this calculation, and the effect of unphysical quark masses on the results could not be quantified. These limitations motivate a follow-up calculation at physical quark masses with improved statistical precision to further benchmark the CG method.

In this work, we perform a high-statistics calculation of the proton isovector unpolarized PDF at physical quark masses on a $(2+1)$-flavor ensemble with $a=0.076$~fm, using $P_z=1.8$ and $2.3$~GeV. To retain statistical precision at these large boosts, we use the kinematically enhanced interpolating operators~\cite{Zhang:2025hyo,Reitinger:2026hta} together with the CG method. The extracted bare matrix elements are renormalized using both the static regularization-independent momentum subtraction (SRI$^\prime$-MOM) scheme developed in Ref.~\cite{Gao:2026hix} and the hybrid scheme~\cite{Ji:2020brr}.
We compare the renormalized matrix elements at short distances with those obtained from the NNPDF4.0 fit~\cite{NNPDF:2021njg} through next-to-leading-order (NLO) perturbative matching and find consistency in both renormalization schemes. In particular, we observe close agreement in the imaginary part of the matrix elements for the first time. We then obtain the quasi-PDF by Fourier transforming the matrix elements with asymptotic extrapolation~\cite{Ji:2026vir}, followed by matching to the light-cone PDF at NLO with next-to-leading-logarithmic (NLL) resummation~\cite{Su:2022fiu,Ding:2024saz}. The final PDF is consistent with our previous calculation~\cite{Gao:2026hix} and again shows good agreement with NNPDF4.0 within statistical and systematic uncertainties in the region of $0.25\lesssim|x|\lesssim0.75$. Together with our previous work, these encouraging results demonstrate the robustness of the CG method for precision PDF calculations and motivate further systematic improvements toward a more stringent benchmark against phenomenology.

The remainder of this paper is organized as follows. Section~\ref{sec:theory} presents the CG framework and our analysis methods. Section~\ref{sec:lattice} describes the lattice setup and the extraction of bare matrix elements. Section~\ref{sec:results} presents the numerical results and comparisons with phenomenological PDFs in coordinate and $x$ space, as well as with earlier lattice calculations. A summary is given in Sec.~\ref{sec:summary}.

\section{\label{sec:theory}Theoretical framework and analysis methods}

\subsection{\label{subsec:cg}Light-cone PDFs from Coulomb-gauge quasi-PDFs}

The unpolarized quark PDF is defined by the light-cone correlation
\begin{equation}
\begin{aligned}
  q(x,\mu) &= \int_{-\infty}^{\infty} \frac{{\rm d}\xi^-}{4\pi}\,
    e^{-ixP^+\xi^-} \\
  &\quad\times \langle P|\bar\psi(\xi^-)\gamma^+
    W(\xi^-,0)\psi(0)|P\rangle_{\mu},
\end{aligned}
\label{eq:lc_pdf}
\end{equation}
where $|P\rangle$ is a hadron state with momentum $P=(P_t,0,0,P_z)$, $P^+=(P_t+P_z)/\sqrt{2}$, and $W(\xi^-,0)$ is the lightlike Wilson line required for gauge invariance~\cite{Collins:2011zzd}. The spin average is implicit. The PDF is understood to be renormalized in the $\overline{\rm MS}$ scheme at the scale $\mu$. For $x>0$, $\bar{q}(x,\mu)$ denotes the antiquark PDF corresponding to $q(x,\mu)$, and we use the convention $q(-x,\mu)=-\bar{q}(x,\mu)$.

The nonzero lightlike separation entering Eq.~\eqref{eq:lc_pdf} cannot be accessed directly from Euclidean correlation functions. LaMET instead relates the light-cone distribution to a quasi-PDF defined from equal-time spatial correlations in a hadron moving at large momentum. In the
present calculation, we choose the equal-time CG correlator as the large-momentum Euclidean observable. For a hadron boosted along the $z$ direction, we define
\begin{equation}
  \widetilde q^{{\rm R}}(x,P_z) = P_z\int_{-\infty}^{\infty} \frac{{\rm d}z}{2\pi}\,e^{-ixP_z z}\,h_q^{{\rm R}}(z,P_z),
\label{eq:cg_quasi_pdf}
\end{equation}
where the corresponding bare correlator is
\begin{equation}
  h_q^{\rm B}(z,P_z,a) = \frac{1}{2P_t} \left\langle P\left| \left.\bar\psi(0)\gamma^t\psi(z\hat z)\right|_{\rm CG}\right|P\right\rangle.
\label{eq:cg_matrix}
\end{equation}
Here $h_q^{\rm B}$ is the bare quantity extracted directly from the lattice data and $h_q^{{\rm R}}$ denotes the scheme-dependent renormalized correlator described in Sec.~\ref{subsec:renormalization}. In the following equations, we suppress the superscript \({\rm R}\) on renormalized correlators and quasi-PDFs when no ambiguity arises. The prefactor $1/(2P_t)$ accounts for the relativistic normalization of the hadron states~\cite{Peskin:1995ev}. The subscript ${\rm CG}$ denotes that the gauge fields are fixed to Coulomb gauge, $\boldsymbol{\nabla}\cdot\boldsymbol A=0$.

The CG correlator in Eq.~\eqref{eq:cg_matrix} constitutes a distinct large-momentum Euclidean observable from the conventional GI quasi-PDF. Its connection to the light-cone PDF follows from their common leading-power limit. Under an infinite boost along the $z$ direction and with an appropriate boundary condition, Coulomb gauge approaches light-cone gauge, in which the lightlike Wilson line reduces to the identity~\cite{Zhao:2023ptv}. At large but finite $P_z$, the CG quasi-PDF therefore admits a LaMET factorization onto the same light-cone PDF, with the dependence on the Euclidean operator construction contained in a perturbatively calculable coefficient and with residual corrections suppressed by inverse powers of $P_z$. The one-loop calculation supports this factorization at that order by showing that the CG quasi-PDF and the light-cone PDF have identical collinear singularities~\cite{Gao:2023lny}. In contrast to the GI quasi-PDF, the CG correlator contains no spatial Wilson line and is consequently free of the associated linear power divergence and linear-renormalon ambiguity~\cite{Braun:2018brg,Zhang:2023bxs,Liu:2023onm,Gao:2023lny,Zhang:2024omt}. This substantially simplifies its renormalization structure.

For the isovector distribution considered here, the coordinate-space matrix element is
\begin{equation}
  h^{\rm B}(z,P_z,a) = \frac{1}{2P_t}\bigl\langle P\bigl|
    \bigl[\bar u(0)\gamma^t u(z\hat z) -\bar d(0)\gamma^t d(z\hat z)\bigr]_{\rm CG}
    \bigr|P\bigr\rangle.
\label{eq:isovector_cg}
\end{equation} 
For either a single flavor or the \(u-d\) combination, according to Hermiticity and translation invariance, the matrix element obeys
\begin{equation}
  h_q(-z,P_z)=h_q^*(z,P_z),
\label{eq:h_symmetry}
\end{equation}
so that $\Re[h_q]$ is even in $z$ and $\Im[h_q]$ is odd. Consequently, the quasi-PDF is real and can be written as
\begin{equation}
\begin{aligned}
  \widetilde q(x,P_z) &= 2P_z\int_0^\infty\frac{{\rm d}z}{2\pi}
    \bigl[\cos(xP_z z)\,\Re h_q(z,P_z) \\
  &\qquad+\sin(xP_z z)\,\Im h_q(z,P_z)\bigr].
\end{aligned}
\label{eq:quasi_reim}
\end{equation}
At finite $P_z$, the quasi-PDF is not restricted to the physical support $|x|\leq1$. For $x>0$, we introduce $\widetilde{\bar q}(x,P_z)\equiv-\widetilde q(-x,P_z)$ and form the valence and full-quark combinations
\begin{equation}
\begin{aligned}
  \widetilde q_v(x,P_z) &\equiv \widetilde q(x,P_z)-\widetilde{\bar q}(x,P_z) = \widetilde q(x,P_z)+\widetilde q(-x,P_z), \\
  \widetilde q_+(x,P_z) &\equiv \widetilde q(x,P_z)+\widetilde{\bar q}(x,P_z) = \widetilde q(x,P_z)-\widetilde q(-x,P_z).
\label{eq:qv_qplus}
\end{aligned}
\end{equation}
Using Eq.~\eqref{eq:quasi_reim}, these combinations are determined separately by the real and imaginary parts of the coordinate-space matrix element,
\begin{equation}
\begin{aligned}
  \widetilde q_v(x,P_z) &= 4P_z\int_0^\infty\frac{{\rm d}z}{2\pi} \cos(xP_z z)\, \Re[h_q(z,P_z)], \\
  \widetilde q_+(x,P_z) &= 4P_z\int_0^\infty\frac{{\rm d}z}{2\pi} \sin(xP_z z)\, \Im[h_q(z,P_z)].
\label{eq:qv_qplus_reim}
\end{aligned}
\end{equation}
For the isovector channel, the corresponding light-cone combinations are
\begin{equation}
  q_v^{u-d} = (u-\bar u)-(d-\bar d), \quad q_+^{u-d} =(u+\bar u)-(d+\bar d).
\label{eq:qv_qplus_lc}
\end{equation} 

\subsection{\label{subsec:renormalization}Renormalization of bare CG quasi-PDF matrix elements}
For the nonlocal CG bilinear, the ultraviolet (UV) renormalization is multiplicative: a quark-field renormalization factor independent of the separation $z$ and the external hadron state relates the bare and renormalized operators,
\begin{equation}
    \left.\bar\psi^{\rm B}(0)\Gamma\psi^{\rm B}(z)\right|_{\rm CG}
    = Z_{\rm R}(a,\mu)
      \left[\bar\psi(0)\Gamma\psi(z)\right]_{{\rm R},{\rm CG}},
    \label{eq:cg_multiplicative}
\end{equation}
where $\Gamma=\gamma^t$ in this work. We use this multiplicative property in two complementary prescriptions. SRI$^\prime$-MOM determines the factor from the equal-time quark propagator and applies it directly to the matrix element at each separation~\cite{Gao:2026hix}. The hybrid prescription cancels the common factor through ratios with a reference matrix element of the same operator~\cite{Ji:2020brr,Gao:2023lny}. Because the two prescriptions differ by finite renormalization conventions, their quasi-PDFs require scheme-specific perturbative matching kernels to target the same $\overline{\rm MS}$ light-cone PDF, as discussed in Sec.~\ref{subsec:matching}. Applying both prescriptions to the same bare matrix elements allows us to probe the residual renormalization-prescription dependence at the lattice spacing and perturbative order used here.

\subsubsection{SRI$^\prime$-MOM Scheme}

We determine $Z_{\rm R}$ from the equal-time CG quark propagator using the SRI$^\prime$-MOM prescription~\cite{Gao:2026hix}, a modification of the conventional RI$^\prime$-MOM~\cite{Martinelli:1994ty} scheme. After conversion to $\overline{\rm MS}$, we denote the resulting factor by $Z_{\overline{\rm MS}}$ and obtain
\begin{equation}
    Z_{\overline{\rm MS}}(a,2~{\rm GeV}) = 0.9114(89),
\end{equation}
for this ensemble. Its determination is given in Appendix~\ref{app:srimom}. The renormalized matrix element is
\begin{equation}
    h^{\overline{\rm MS}}(z,P_z,\mu)=\frac{h^{\rm B}(z,P_z,a)}{Z_{\overline{\rm MS}}(a,\mu)}.
    \label{eq:sri_h}
\end{equation}
At a fixed renormalization scale, the separation-independent factor commutes with the linear Fourier transform and can therefore be applied either to the coordinate-space matrix element or to the momentum-space quasi-PDF~\cite{Grebe:2026qmt}. We apply it uniformly to all separations in coordinate space before the Fourier transform and do not impose an additional unit normalization at $z=0$. The resulting difference from the conventional local-current normalization contributes only through a contact term under the Fourier transform, yielding an \(\mathcal{O}(a)\) discretization effect in \(x\) space that was found to be negligible in Ref.~\cite{Grebe:2026qmt}.

\subsubsection{Hybrid Scheme}
We choose a zero-momentum proton as the reference state in this work. At short distances, taking a ratio cancels the common UV factor and can reduce effects shared by the numerator and denominator. Beyond a transition distance $\zs$ which satisfies $a\ll z_s \ll 1/\Lambda_{\rm QCD}$, the reference denominator is held fixed so that its long-distance separation dependence does not enter the result:
\begin{equation}
\begin{aligned}
    h^{\rm hyb}(z,\zs,P_z)
    &=N(P_z)h^{\rm B}(z,P_z,a) \\
    &\quad\times\left[
      \frac{\theta(\zs-|z|)}{h^{\rm B}(z,0,a)}
      +\frac{\theta(|z|-\zs)}{h^{\rm B}(\zs,0,a)}
    \right].
\end{aligned}
    \label{eq:hybrid_cg}
\end{equation}
Here $N(P_z)=h^{\rm B}(0,0,a)/h^{\rm B}(0,P_z,a)$ imposes $h^{\rm hyb}(0,\zs,P_z)=1$, the local normalization associated with the isovector valence-number sum rule. Unlike the SRI$^\prime$-MOM factor, the reference ratio depends on $z$ and must be formed before the Fourier transform. We use $\zs=2a=0.152$~fm in the main analysis; the dependence on this choice is examined in Appendix~\ref{app:hybrid_comparison}.

\subsection{\label{subsec:nla_and_ft}Fourier transform from renormalized matrix elements to quasi-PDFs}

The long-distance behavior of the coordinate-space correlator is an important ingredient in the Fourier reconstruction of the quasi-PDF. In a theory with a finite mass gap, spacelike hadronic correlations are expected to be exponentially suppressed at asymptotically large separations, with subleading corrections organized in inverse powers of the separation~\cite{Ji:2026vir}. In practice, the need for an explicit long-distance continuation depends on the observed behavior of the correlator. For the CG matrix elements considered here, the signal can already decay to values compatible with zero within the accessible lattice range. A discrete Fourier transform is then generally sufficient. Residual statistical fluctuations at the largest separations can nevertheless induce additional sensitivity in the reconstructed distribution, most notably at small $x$, where long-distance correlations become increasingly relevant. In such cases, the field-theory-motivated asymptotic form is used primarily to stabilize the Fourier reconstruction rather than to supply a sizable unresolved physical tail.

We therefore constrain the unresolved long-distance contribution using an asymptotic form motivated by the spectral structure of spacelike Euclidean correlators~\cite{Chen:2025cxr,Ji:2026vir,Gao:2026hix}. The large-$|z|$ behavior is governed by an exponential associated with the lowest relevant mass gap, multiplied by a systematic expansion in inverse powers of the separation. Retaining the first subleading term, we employ the next-to-leading-asymptotic (NLA)-inspired parametrization
\begin{equation}
    h^{\text{NLA}}(z,P_z) =\frac{e^{-m|z|}}{|\lambda|^n}
      \biggl[A_2 e^{i\,\operatorname{sgn}(\lambda)\phi_2}+\frac{A'_2}{|\lambda|}
       e^{i\,\operatorname{sgn}(\lambda)\phi'_2}\biggr],
\label{eq:extrapolation_nla}
\end{equation}
where $\lambda=zP_z$ and $\{A_2,A'_2,\phi_2,\phi'_2,m,n\}$ are determined from the lattice data. The exponential factor captures the expected mass-gap suppression, while the overall factor $1/|\lambda|^n$ parametrizes the algebraic decay of the CG correlator, whose precise form has not yet been established theoretically, since the asymptotic form was derived for the GI formulation.

For each momentum and renormalization prescription, the real and imaginary parts of the correlator are fitted simultaneously over the chosen long-distance window, with a common set of asymptotic parameters. The Fourier reconstruction then retains the lattice data where they remain statistically resolved and replaces only the long-distance region by the fitted asymptotic continuation,
\begin{equation}
    h^{\rm R,ext}(z,P_z)=\begin{cases}h^{\rm R}(z,P_z),& |z|<z_{\rm extrap}, \\[2mm]h^{\rm NLA}(z,P_z), & |z|\ge z_{\rm extrap}.\end{cases}
\label{eq:piecewise_m}
\end{equation}
The switching point $z_{\rm extrap}$ is chosen within the region where the lattice data are still sufficiently precise to constrain the asymptotic form, so that the transition does not rely on an uncontrolled continuation from noise-dominated data.

The quasi-PDFs in Eq.~\eqref{eq:qv_qplus_reim} are subsequently obtained by Fourier transforming the extended correlator. The measured region is integrated directly, while the asymptotic continuation supplies the contribution over $z_{\rm extrap}\le |z|<\infty$. The fit ranges, parameter choices, and resulting long-distance extrapolations are discussed in Sec.~\ref{subsec:results_extrap_ft}.

\subsection{\label{subsec:matching}Perturbative matching with renormalization-group resummation}

The quasi-PDF and the light-cone PDF share the same long-distance collinear physics but differ in their short-distance structure. LaMET accounts for these differences through a perturbatively calculable matching coefficient, relating a quasi-PDF renormalized in a scheme ${\rm R}$ to the $\overline{\rm MS}$ light-cone PDF~\cite{Ji:2013dva,Ji:2014gla,Izubuchi:2018srq,Yao:2022vtp,Ji:2020ect,Gao:2023lny}:
\begin{equation}
\begin{aligned}
    q^{\overline{\rm MS}}(x,\mu)
    &= \int_{-\infty}^{\infty}\frac{{\rm d}y}{|y|}\,
       C^{\rm R}\!\left(\frac{x}{y},\frac{\mu}{p_z}\right)
       \widetilde q^{\rm R}(y,P_z) \\
    &\quad+ \mathcal O\!\left(
      \frac{\Lambda_{\rm QCD}^2}{x^2P_z^2},
      \frac{\Lambda_{\rm QCD}^2}{(1-|x|)^2P_z^2}
    \right),
\end{aligned}
    \label{eq:factorization}
\end{equation}
Here $p_z\equiv|y|P_z$. The choice ${\rm R}=\overline{\rm MS}$ denotes the SRI$^\prime$-MOM analysis after conversion to $\overline{\rm MS}$, whereas ${\rm R}={\rm hyb}$ denotes the hybrid prescription. For ${\rm R}={\rm hyb}$, the dependence of both $\widetilde q^{\rm R}$ and $C^{\rm R}$ on $\zs$ is implicit. Although the quasi-PDFs and matching coefficients depend on the renormalization prescription, both analyses target the same light-cone PDF. The matching coefficient is known at NLO for all spin structures~\cite{Gao:2023lny,Gao:2026hix}. The explicit one-loop kernels are given in Appendix~\ref{app:matching}.

The perturbative matching coefficients contain collinear logarithms governed by the partonic momentum rather than the hadron momentum alone. For a target momentum fraction $x$, the relevant external scale is $2|x|P_z$, which can differ substantially from the final factorization scale $\mu$~\cite{Su:2022fiu,Ding:2024saz}. To resum these logarithms, we implement renormalization-group resummation (RGR) by performing the matching at an initial scale $\mu_0=2\kappa|x|P_z$, with $\kappa$ being a constant, and then evolving the matched distribution to $\mu=2$~GeV in the $\overline{\rm MS}$ scheme using nonsinglet Dokshitzer-Gribov-Lipatov-Altarelli-Parisi (DGLAP) evolution. Combining NLO matching with NLL evolution gives the NLO+NLL results presented below. The isovector distributions evolve without mixing with the gluon PDF. We use $\kappa=1$ as the central scale and vary $\kappa\in\{1/\sqrt{2},1,\sqrt{2}\}$ to assess the residual perturbative scale dependence. The evolution equations and implementation are described in Appendix~\ref{app:rgr}.

RGR improves the treatment of collinear logarithms but does not remove the finite-momentum power corrections in Eq.~\eqref{eq:factorization}. These corrections are enhanced when either $|x|P_z$ or $(1-|x|)P_z$ approaches hadronic scales. At small $|x|$, the decreasing initial scale also leads to a growing coupling and eventually limits perturbative control. As $|x|\to1$, threshold logarithms become important~\cite{Gao:2021hxl,Ji:2023pba,Ji:2024hit}; their resummation is not included in the present analysis~\cite{Gao:2026hix}. These limitations motivate restricting quantitative comparisons to intermediate $|x|$, as discussed in Sec.~\ref{subsec:pdf_results}.

\section{\label{sec:lattice}Lattice setup and matrix-element extraction}

\subsection{Lattice ensemble and measurement setup}

We use a $2+1$-flavor gauge ensemble generated by the HotQCD Collaboration~\cite{Bazavov:2019www} with the highly improved staggered quark (HISQ) action~\cite{Follana:2006rc} and sea quark masses tuned to their physical values. The lattice dimensions are $L_{\rm s}^3\times L_{\rm t}=64^3\times64$, with a lattice spacing of $a=0.076$~fm. For the valence quarks, we employ the tree-level tadpole-improved clover-Wilson action~\cite{Sheikholeslami:1985ij,Lepage:1992xa} with $c_{\rm SW}=1.0372$, and tune the quark mass to yield $m_\pi\simeq140$~MeV. For the CG measurements, we fix each gauge configuration independently on every Euclidean time slice by finding a gauge transformation $\Omega$ that minimizes
\begin{equation}
  F_t[U^\Omega]=-\frac{1}{9V_s} \sum_{\mathbf{x}}\sum_{i=1}^{3}
  \operatorname{ReTr} U_i^\Omega(t,\mathbf{x}),
\end{equation}
where $V_s=L_{\rm s}^3$ and $U_i^\Omega$ denotes the gauge-transformed spatial links. We perform the CG fixing using the \href{https://github.com/RJHudspith/GLU}{Gauge Link Utility (GLU)} code~\cite{Hudspith:2014oja}, with a convergence tolerance of $10^{-14}$. The minimization may nevertheless converge to different local minima of $F_t$ on the same gauge orbit (Gribov copies)~\cite{Gribov:1977wm,Singer:1978dk}, which can in principle affect gauge-dependent correlators. We retain the first converged CG copy obtained by the gauge-fixing procedure. Previous CG studies suggest that Gribov-copy effects are subdominant to statistical uncertainties in the observables examined~\cite{Gao:2024fbh}. The links entering the valence action are subsequently subjected to one step of hypercubic (HYP) smearing~\cite{Hasenfratz:2001hp}.

We compute matrix elements at on-axis momenta $\mathbf{n}=(0,0,n_z)$, with $P_z=2\pi n_z/(aL_{\rm s})$. The choices $n_z=7$ and $9$ give $P_z=1.8$ and $2.3$~GeV; the $n_z=0$ matrix elements provide the reference for hybrid renormalization in Eq.~\eqref{eq:hybrid_cg}. To improve precision at large boosts and spatial quark separations, we use momentum-smeared Gaussian quark sources~\cite{Bali:2016lva}, a kinematically enhanced nucleon interpolating operator~\cite{Zhang:2025hyo,Reitinger:2026hta}, and CG quark bilinears~\cite{Gao:2023lny}. We construct Gaussian-smeared quark sources with smearing momentum $K_z=\zeta P_z$ to improve the overlap of the resulting proton interpolating operators with boosted states. 

The enhanced interpolating operator has been shown to improve the signal-to-noise ratio (SNR) at large $P_z$. The CG bilinear, defined on CG-fixed configurations without a spatial Wilson line, avoids a source of large-separation SNR degradation in the bare matrix elements and the associated Wilson-line self-energy renormalization factor, which can amplify absolute statistical uncertainties~\cite{Gao:2023lny,Gao:2026hix}. We use four source-sink separations, $\ts/a\in\{6,8,10,12\}$, to assess excited-state contamination (ESC). We obtain all quark propagators through high-precision Dirac-operator inversions with a solver stopping tolerance of \(10^{-10}\). Measurements are performed on 700 gauge configurations. The smearing parameters and measurement statistics are summarized in Table~\ref{tab:lattice_setup}.

\begin{table}[htpb]
    \centering
    \caption{Summary of the lattice measurement setup and statistics. $P_z$ denotes the physical proton momentum along the $z$-direction, corresponding to the discrete lattice momentum $n_z$. The momentum smearing parameter is denoted by $\zeta$. The matrix elements are extracted using four source-sink time separations $\ts$. \(N_{\rm src}\) denotes the number of source positions per gauge configuration for each source-sink separation. The total number of measurements is $N_{\rm meas} = N_{{\rm cfg}} \times N_{\rm src}$ with $N_{\rm cfg}=700$.}
    \label{tab:lattice_setup}
    \begin{tabular}{c|c|c|c|c|c}
        \hline\hline
        $P_z$ [GeV] & $n_z$ & $\zeta$ & $\ts/a$ & $N_{\rm src}$ & $N_{\rm meas}$ \\
        \hline
        \multirow{2}*{$0$} & \multirow{2}*{$0$} & \multirow{2}*{0} & $6, 8$ & 16 & 11,200 \\
         & & & $10, 12$ & 32 & 22,400 \\
        \hline
        \multirow{3}*{$1.8$} & \multirow{3}*{$7$} & \multirow{3}*{3/7} & $6, 8$ & 32 & 22,400 \\
        & & & $10$ & 128 & 89,600 \\
        & & & $12$ & 512 & 358,400 \\
        \hline
        \multirow{3}*{$2.3$} & \multirow{3}*{$9$} & \multirow{3}*{1/3} & $6, 8$ & 32 & 22,400 \\
        & & & $10$ & 512 & 358,400 \\
        & & & $12$ & 1024 & 716,800 \\
        \hline\hline
    \end{tabular}
\end{table}

\subsection{\label{subsec:me_extraction}Ground-state matrix-element extraction}

The bare spatial matrix elements $h^{\rm B}(z, P_z, a)$ are extracted by evaluating the two-point and three-point correlation functions. To create and annihilate the proton states on the lattice, we employ the interpolating operator defined as
\begin{equation}
    N_{\alpha}(\mathbf{x}, t) = \epsilon_{abc}\; u_{a,\alpha}(\mathbf{x}, t) \left[ u_b^T(\mathbf{x}, t) C\gamma_t\gamma_5 d_c(\mathbf{x}, t) \right],
\end{equation}
where the subscripts $\{a, b, c\}$ denote the color indices contracted by the totally antisymmetric Levi-Civita tensor $\epsilon_{abc}$, $\alpha$ is the Dirac spinor index, and $C=\gamma^t\gamma^y$ is the charge conjugation matrix. This choice replaces the conventional scalar diquark $C\gamma_5$ with the kinematically enhanced structure $C\gamma_t\gamma_5$,
which has been shown to improve the SNR at large boosts~\cite{Zhang:2025hyo} and is therefore well suited for the large-$P_z$ program pursued here.

In the forward limit relevant for PDFs, the initial and final momenta are identical ($\mathbf{P}^i = \mathbf{P}^f = \mathbf{P}$). The unpolarized two-point correlation function is constructed as
\begin{equation}
    C_{{\rm 2pt}}(\mathbf{P}, \ts) = \sum_{\mathbf{x}} e^{-i\mathbf{P}\cdot\mathbf{x}} \; \mathcal{P}^{+}_{\alpha\beta} \langle N_\beta(\mathbf{x}, \ts) \bar{N}_\alpha(\mathbf{0}, 0) \rangle, \label{eq:c2pt}
\end{equation}
where $\mathcal{P}^{+} = (1+\gamma^t)/2$ is the positive-parity projector; no additional spin projection is applied.

To evaluate the quasi-PDF matrix elements, the corresponding three-point correlation functions are computed by inserting the nonlocal CG quark bilinear operator $O^{\gamma^t}(z) = \bar{q}(0)\gamma^t q(z)$ at an intermediate time $\tau$. The three-point function is given by
\begin{equation}
\begin{aligned}
    C_{{\rm 3pt}}(\mathbf{P}, \tau, \ts; z)
    &= \sum_{\mathbf{x},\mathbf{y}} e^{-i\mathbf{P}\cdot\mathbf{x}}
       \mathcal{P}^{+}_{\alpha\beta} \\
    &\quad\times\bigl\langle N_\beta(\mathbf{x}, \ts)
       O^{\gamma^t}(\mathbf{y}, \tau; z)
       \bar N_\alpha(\mathbf{0},0)\bigr\rangle.
\end{aligned}
\label{eq:c3pt}
\end{equation}

By inserting a complete set of intermediate hadronic states, the temporal dependence of the correlation functions can be expressed via their spectral decompositions:
\begin{align}
    C_{{\rm 2pt}}(\mathbf{P}, \ts) &= \sum_{n=0}^{\infty} |A_n|^2 e^{-E_n \ts}, \label{eq:c2pt_spec} \\
    C_{{\rm 3pt}}(\mathbf{P}, \tau, \ts; z) &= \sum_{m,n=0}^{\infty} A_m h_{mn} A_n^* e^{-E_m(\ts-\tau)} e^{-E_n \tau}, \label{eq:c3pt_spec}
\end{align}
Here, $E_n$ represents the energy of the $n$-th state. The overlap amplitude $A_n=\langle \Omega|N \cdot \mathcal{P}^{+} |n, \mathbf{P} \rangle / \sqrt{2E_n}$ quantifies the transition coupling between the vacuum and the $n$-th hadronic state under the unpolarized projection operator. The term $h_{mn} = \frac{\langle m, \mathbf{P} | O^{\gamma^t} | n, \mathbf{P} \rangle}{2\sqrt{E_mE_n}}$ denotes the transition matrix element. In particular, $h_{00}$ is the dimensionless ground-state bare matrix element, $h^{\rm B}$, defined with the $1/(2E_0)$ normalization in Eq.~\eqref{eq:isovector_cg}.

To isolate the ground-state contribution, we construct the standard ratio:
\begin{equation}
    R(\mathbf{P}; \tau, \ts; z) \equiv \frac{C_{{\rm 3pt}}(\mathbf{P}, \tau, \ts; z)}{C_{{\rm 2pt}}(\mathbf{P}, \ts)}. \label{eq:ratio}
\end{equation}
In the asymptotic limit where both the source-insertion separation $\tau$ and the insertion-sink separation $(\ts - \tau)$ are sufficiently large, the contributions from excited states decay exponentially faster than the ground-state contribution. Consequently, all higher-order terms vanish, and the ratio approaches the normalized bare matrix element: $\lim_{\tau, (\ts-\tau) \to \infty} R = h^{\rm B}$.

In practice, the asymptotic limit cannot be reached directly because the SNR of the correlation functions deteriorates with increasing $\ts$, while data at finite $\ts$ can receive excited-state contamination (ESC). We therefore extract the ground-state bare matrix element $h^{\rm B}$ using two complementary methods: a two-state fit that explicitly models the lowest excited state and an exponential summation fit that suppresses the leading transition contributions~\cite{Ding:2024lfj,Gao:2020ito}. To incorporate the residual dependence on the extraction method, we combine their bootstrap distributions with equal weight and propagate the resulting samples through all subsequent analysis steps. The fit forms, data selection, and resampling procedure are detailed in Appendix~\ref{app:me_extraction}. Unless noted otherwise, central values of our lattice results are medians, while error bars and bands denote pointwise central percentile intervals containing approximately $68\%$ of the corresponding sample distributions.

\section{\label{sec:results}Numerical results}
\subsection{\label{subsec:results_bare_me}Bare matrix elements}

\begin{figure}[!tp]
    \centering
    \includegraphics[width=0.9\linewidth]{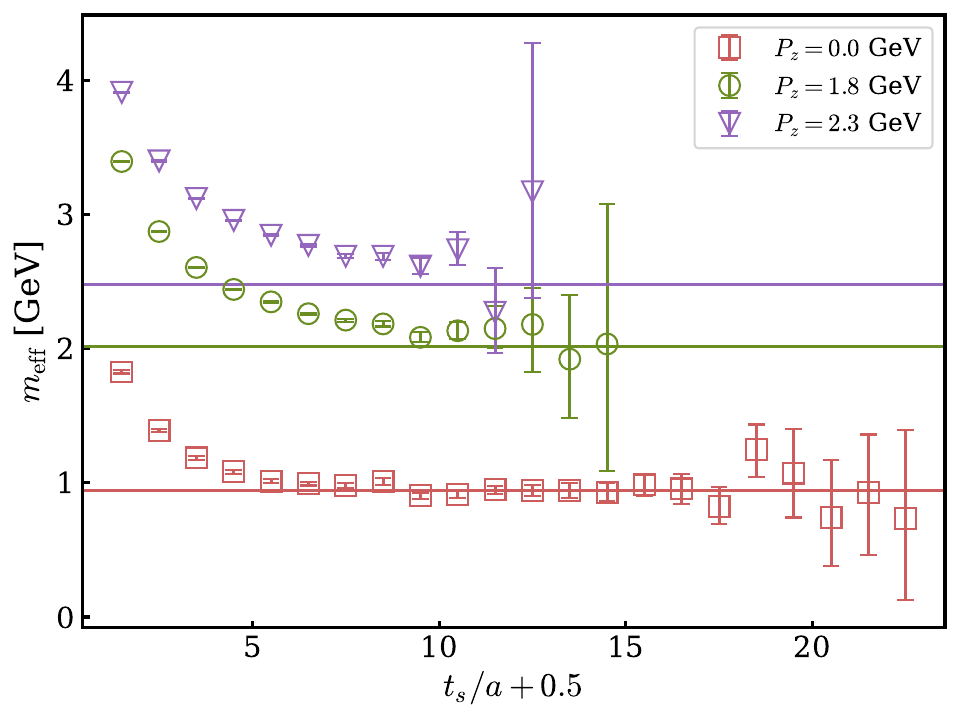}
    \caption{\label{fig:effective_mass} Effective mass $m_{\rm eff}$ plotted at $\ts/a+1/2$ for $P_z \in \{0, 1.8, 2.3\}$~GeV. The colored horizontal lines show the dispersion-relation estimates.}
\end{figure}

Before extracting the bare matrix elements, we first inspect the effective masses obtained from the two-point correlation functions. Figure~\ref{fig:effective_mass} shows the effective mass as a function of $\ts$ for the three momenta $P_z \in \{0, 1.8, 2.3\}$~GeV. The colored lines show the dispersion relation $E(P_z)=\sqrt{m_{\rm N}^2+P_z^2}$ evaluated with a fixed reference proton mass of $m_{\rm N}=0.94$~GeV. At all three momenta, the effective mass tends to plateau around the corresponding predictions when $\ts$ is large.

\begin{figure}[!tp]
    \centering
    \includegraphics[width=0.9\linewidth]{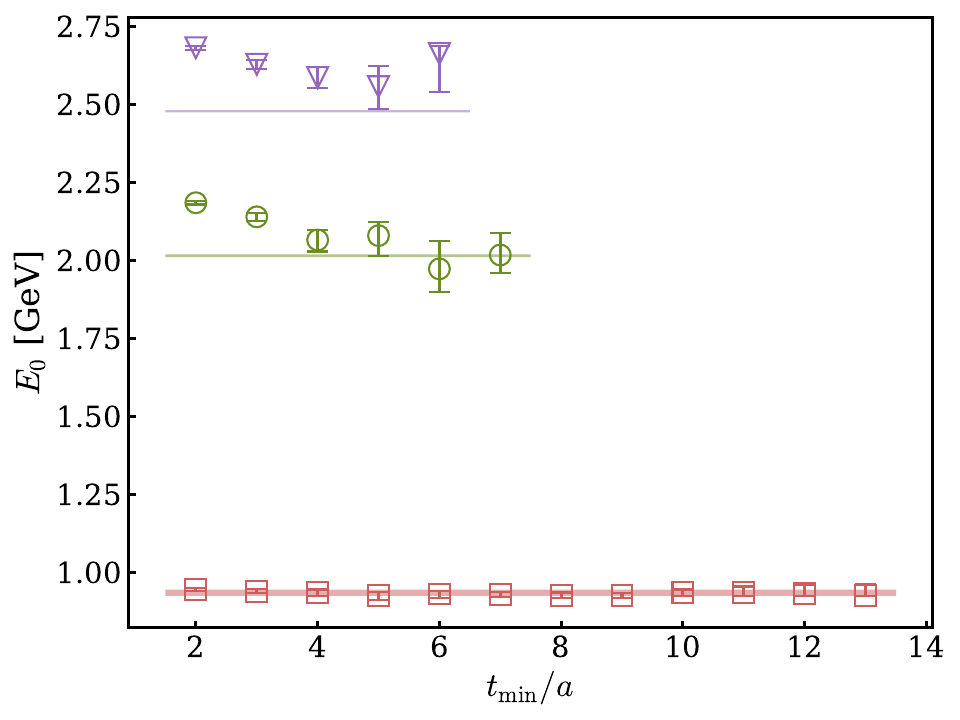}\\[-0.5ex]
    \includegraphics[width=0.9\linewidth]{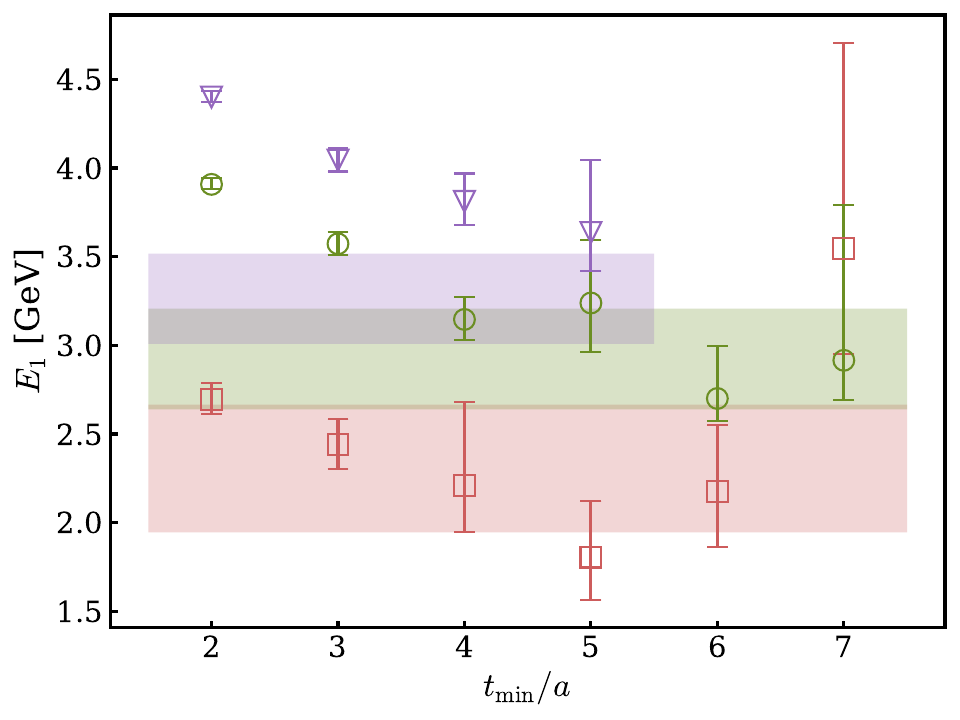}
    \caption{\label{fig:energies} Ground-state energy $E_0$ (top) and excited-state energy $E_1$ (bottom) obtained from two-state fits with a prior on $E_0$, shown as functions of the lower fit bound $t_{\rm min}/a$. The colored bands show the dispersion estimates obtained from the corresponding rest-frame bootstrap samples at $t_{\rm min}/a=4$.}
\end{figure}

We determine the ground- and excited-state energies from two-state fits with a prior on $E_0$. The resulting $E_0$ and $E_1$ are shown as functions of the lower fit bound $t_{\rm min}/a$ in Fig.~\ref{fig:energies}. For $n_z=0$, $7$, and $9$, we impose Gaussian priors on $aE_0$ with $(\mu,\sigma)=(0.357,0.0357)$, $(0.795,0.0795)$, and $(0.996,0.0996)$, respectively. The central values are obtained from standard two-state fits, and the prior widths are set to $10\%$ of the corresponding central values. At zero momentum, the fit with $t_{\rm min}/a=4$ yields
\begin{equation}
\begin{aligned}
    m_{\rm N}\equiv E_0(0)&=0.936(10)~{\rm GeV},\\
    E_1(0)&=2.21^{+0.46}_{-0.27}~{\rm GeV}.
\end{aligned}
\end{equation}
For each level $n=0,1$, the colored bands show the dispersion estimates $E_n(P_z)=\sqrt{E_n^2(0)+P_z^2}$, evaluated sample by sample using this rest-frame fit at $t_{\rm min}/a = 4$. The ground-state energy shows weaker fit-window dependence at larger $t_{\rm min}$, while the excited-state energy remains more sensitive to the fit window and has larger uncertainties. We adopt the results at $t_{\rm min}/a=\{4,5,5\}$ for $P_z=\{0,1.8,2.3\}$~GeV, respectively, as inputs throughout the ratio analysis below.

\begin{figure}[!tbp]
    \centering
    \includegraphics[width=0.98\linewidth]{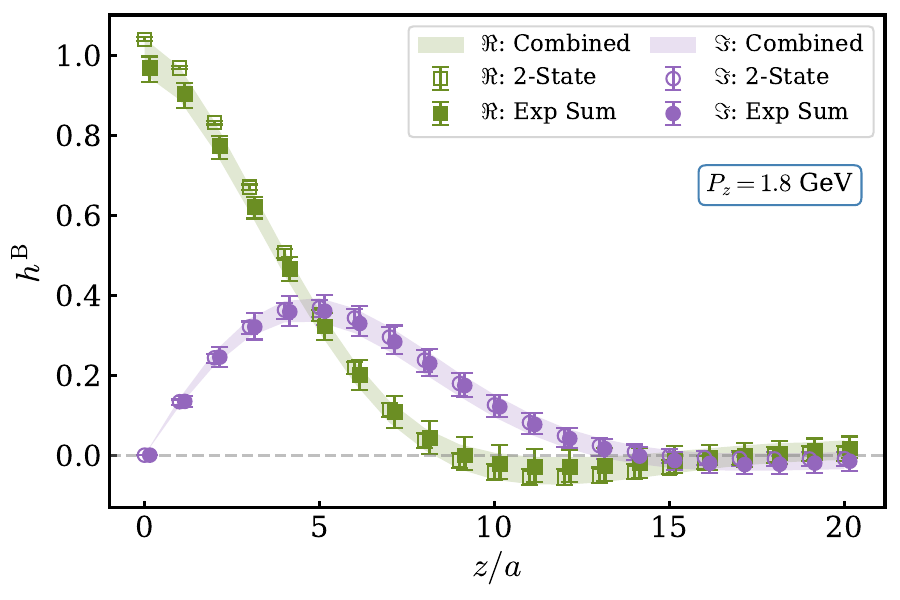}\\[-0.5ex]
    \includegraphics[width=0.98\linewidth]{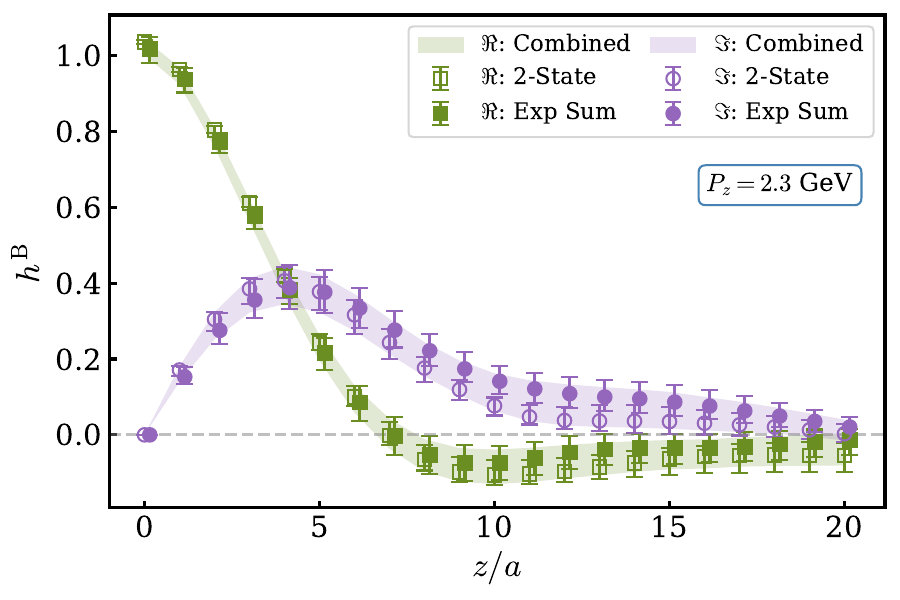}
    \caption{\label{fig:bare_me} Isovector bare matrix elements $h^{{\rm B}}$ as functions of $z/a$, at $P_z=1.8$~GeV (top) and $2.3$~GeV (bottom). Each panel displays the real part as green squares and the imaginary part as purple circles. Open and filled symbols denote the two-state and exponential-summation extractions, respectively. The green and purple bands labeled ``Combined'' show the central $68\%$ intervals of the combined bootstrap distributions.}
\end{figure}

Following Sec.~\ref{subsec:me_extraction}, we extract the ground-state bare matrix elements from the ratio $R(\mathbf{P};\tau,\ts;z)$ using the two complementary procedures summarized there and detailed in Appendix~\ref{app:me_extraction}, which also presents representative ratio fits. Since both procedures are applied to the same data and bootstrap ensembles, their difference reflects the extraction-method dependence. Figure~\ref{fig:bare_me} shows the resulting isovector bare matrix elements as functions of quark separations $z$ for both momenta, with the real and imaginary parts displayed together in each panel. At both momenta, the real part decreases from its local value, crosses zero, and develops a shallow negative minimum, while the imaginary part rises from zero, peaks near $z/a\approx4\text{--}5$, and subsequently decreases. The real part decays faster and reaches its zero crossing at a smaller $z/a$ for the larger momentum.

\begin{figure*}[!tp]
    \centering
    \includegraphics[width=0.47\linewidth]{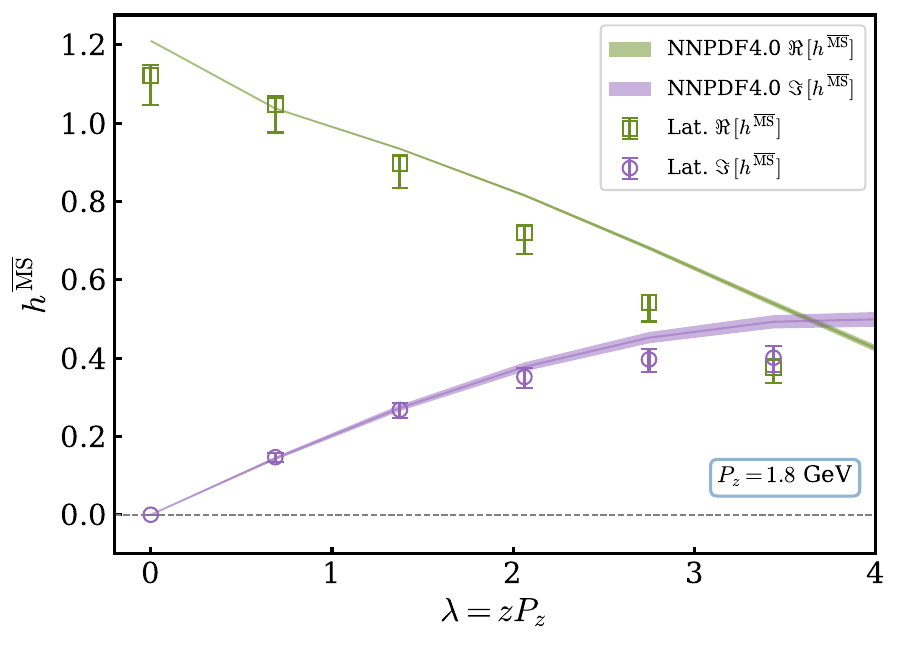}
    \includegraphics[width=0.47\linewidth]{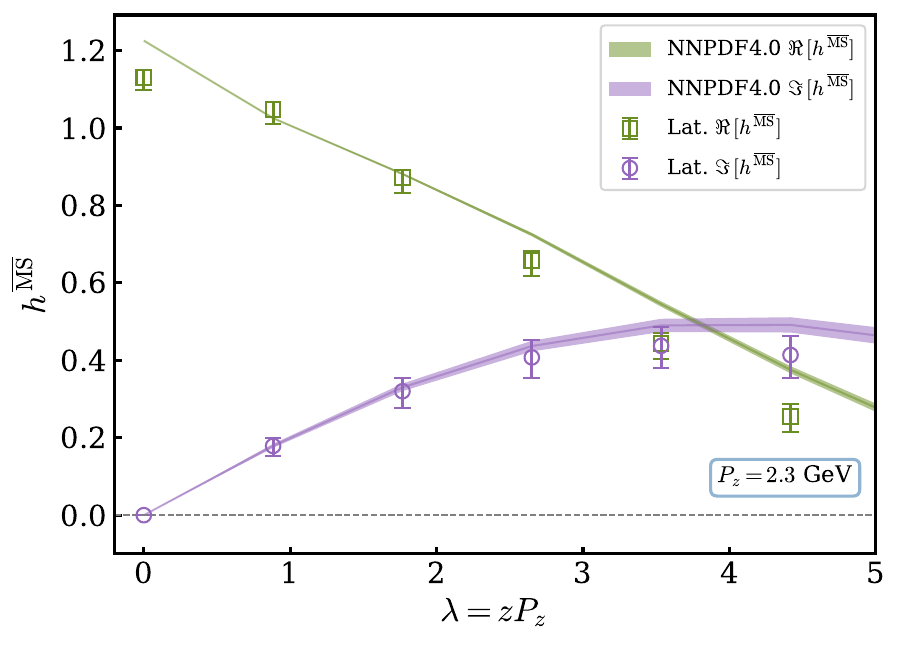} \\
    \includegraphics[width=0.47\linewidth]{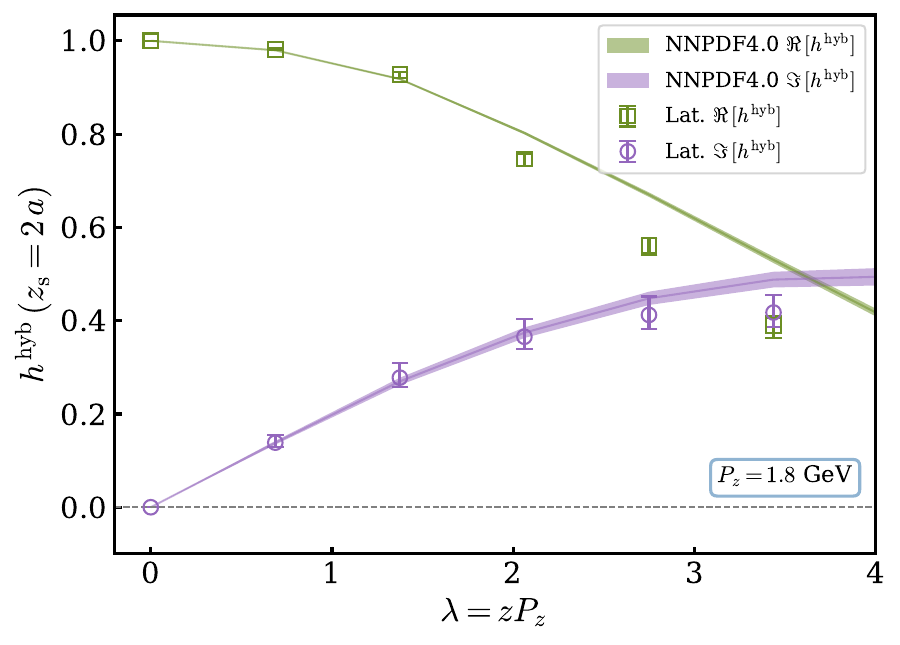}
    \includegraphics[width=0.47\linewidth]{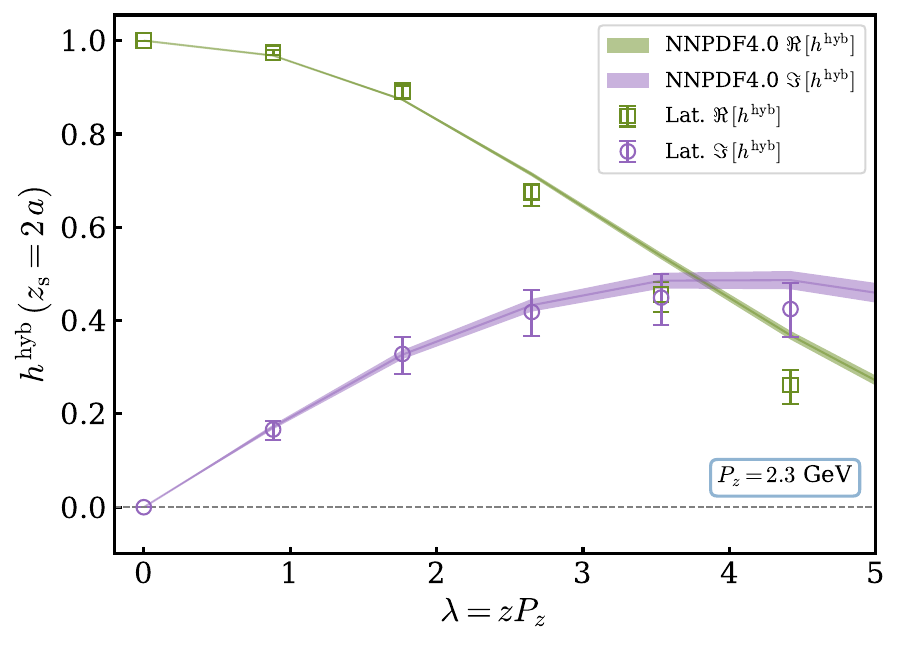}
    \caption{Renormalized isovector matrix elements compared with coordinate-space predictions reconstructed from the NNPDF4.0 PDFs using fixed-order NLO CG matching. Top: SRI$^\prime$-MOM results converted to the $\overline{\rm MS}$ scheme at $\mu=2$~GeV; bottom: hybrid results with $\zs=2a=0.152$~fm. Left and right panels correspond to $P_z=1.8$ and $2.3$~GeV, respectively. Green squares and purple circles show the real and imaginary parts, respectively, while the corresponding bands denote the NNPDF4.0-based predictions.}
    \label{fig:forward_me_comparison}
\end{figure*}

\subsection{Renormalized matrix elements and coordinate-space benchmark}
\label{subsec:renorm_coordinate}

We renormalize the bare matrix elements obtained above using the two prescriptions described in Sec.~\ref{subsec:renormalization}: SRI$^\prime$-MOM followed by perturbative conversion to the $\overline{\rm MS}$ scheme at $\mu=2$~GeV, with $Z_{\overline{\rm MS}}(a,2~{\rm GeV})=0.9114(89)$, and hybrid renormalization with $\zs=2a=0.152$~fm. The resulting renormalized matrix elements can be compared directly with coordinate-space predictions reconstructed from phenomenological PDFs. This comparison is performed directly on the Euclidean observables entering the subsequent quasi-PDF reconstruction. Consequently, it provides a natural stage to assess renormalization and finite-momentum factorization before extrapolating or Fourier transforming the lattice data. The comparison is particularly informative at short distances, where nonperturbative corrections to the leading-power factorization are suppressed by $z^2\Lambda_{\rm QCD}^2$~\cite{Gao:2023lny}.

To construct the phenomenological expectation, we start from the NNPDF4.0 isovector PDF~\cite{NNPDF:2021njg} at $\mu=2$~GeV and apply the forward NLO CG matching kernel appropriate to each renormalization prescription. Transforming the resulting quasi-PDF back to coordinate space gives
\begin{equation}
    h_{\rm pred}^{\rm R}(z,P_z) = \int_{-\infty}^{\infty}{\rm d}x\, e^{ixP_z z} \left[ C^{\rm R} \otimes q_{\rm NNPDF}^{\overline{\rm MS}} \right](x),
    \label{eq:forward_me_reconstruction}
\end{equation}
where ${\rm R}=\overline{\rm MS}$ or ${\rm hyb}$ specifies the renormalization scheme of the reconstructed matrix element. The dependence of $C^{\rm hyb}$ on $\zs$ is implicit; the corresponding comparisons for $\zs=3a$ and $4a$ are presented in Appendix~\ref{app:hybrid_comparison}.

Figure~\ref{fig:forward_me_comparison} compares the lattice matrix elements with the NNPDF4.0-based predictions at both momenta and for both renormalization prescriptions. At short distances, both the real and imaginary parts are compatible within the current uncertainties, with particularly close agreement in the imaginary part observed for the first time in a lattice calculation of the proton isovector unpolarized PDF. The origin of this close agreement may lie in the use of kinematically enhanced interpolating operators, physical quark masses, or the CG method, which can be clarified through controlled comparisons in future studies. Moreover, as the imaginary part of the matrix element has been found to be more sensitive to excited-state contamination in Refs.~\cite{Alexandrou:2019lfo,Gao:2026hix}, it is important to verify that this effect is fully under control through further examination. The robustness of this agreement will also be tested at additional lattice spacings and momenta, together with higher-order matching. After all, at the present precision, the short-distance agreement supports the consistency of the directly computed Euclidean matrix elements with phenomenological PDFs through the NLO CG matching relation. Therefore, the coordinate-space benchmark complements the comparison of the resulting $x$-space light-cone PDFs with global analyses presented below.

\begin{figure*}[!tp]
    \centering
    \includegraphics[width=0.47\linewidth]{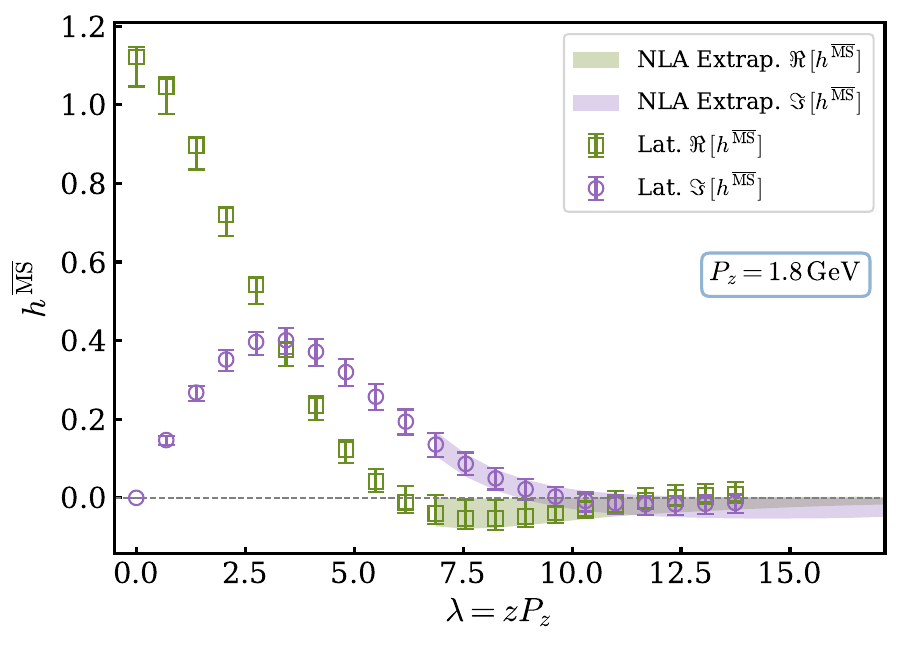}
    \includegraphics[width=0.47\linewidth]{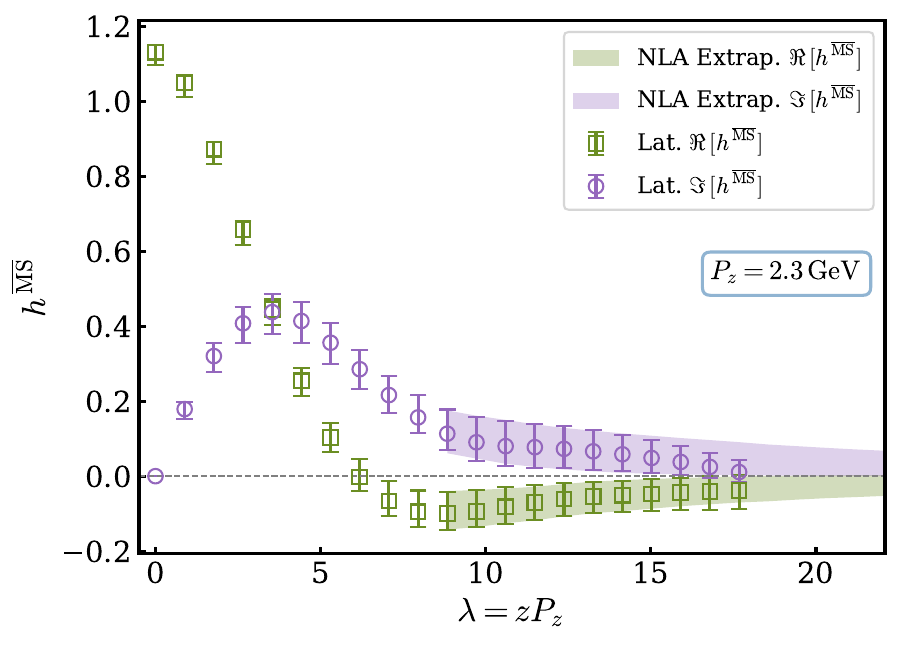} \\
    \includegraphics[width=0.47\linewidth]{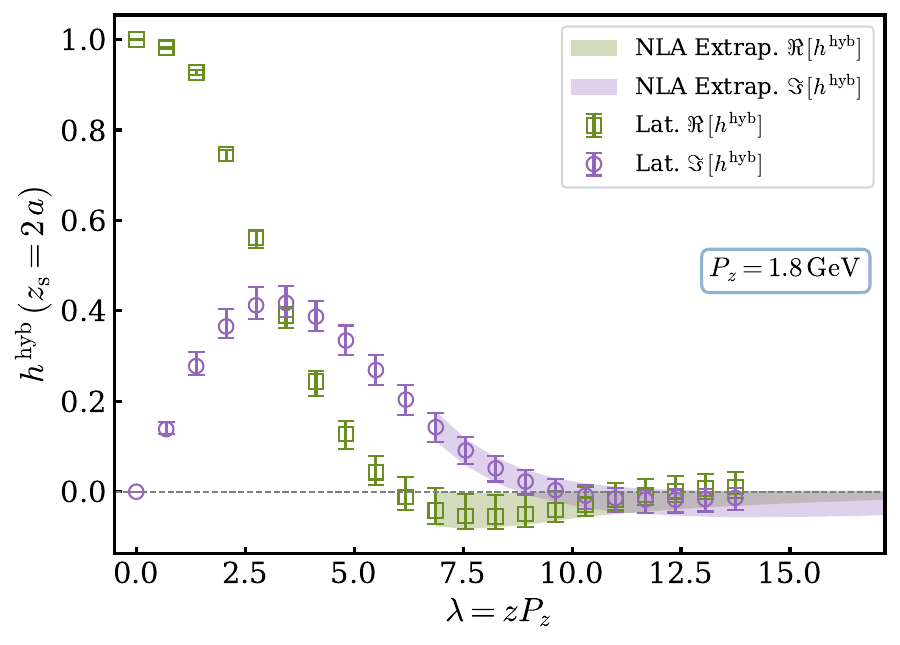}
    \includegraphics[width=0.47\linewidth]{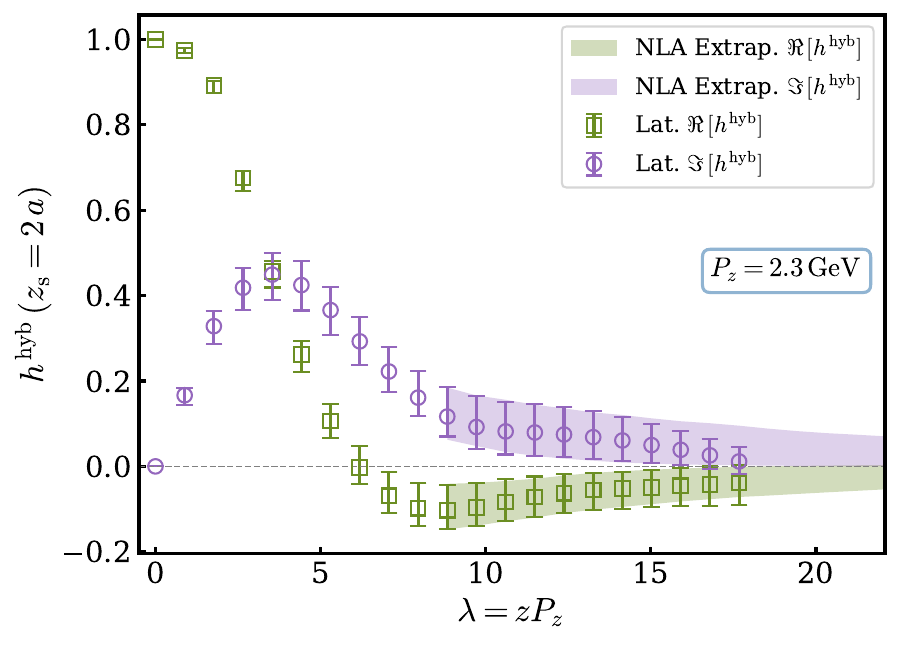}
    \caption{Renormalized isovector matrix elements and their long-distance continuations as functions of $\lambda=zP_z$. Top: SRI$^\prime$-MOM results converted to the $\overline{\rm MS}$ scheme at $\mu=2$~GeV; bottom: hybrid results with $\zs=2a=0.152$~fm. Left and right panels correspond to $P_z=1.8$ and $2.3$~GeV, respectively. Green squares and purple circles denote the real and imaginary parts, respectively. The corresponding bands show the NLA continuations used for the long-distance region.}
    \label{fig:renorm_extrap}
\end{figure*}

\subsection{Quasi-PDF construction}
\label{subsec:results_extrap_ft}

Having examined the renormalized matrix elements in the region directly constrained by the lattice data, we next address their long-distance behavior and the Fourier transform to quasi-PDFs. An explicit long-distance continuation is not intrinsically required if the correlator has already decayed to values compatible with zero and the Fourier transform is stable against variations of the truncation range. Residual fluctuations at the largest separations can, nevertheless, introduce additional sensitivity into the transform, particularly toward small $x$. We therefore use the field-theory-motivated NLA form introduced in Sec.~\ref{subsec:nla_and_ft} to control the long-distance contribution when performing the Fourier transform.

For each momentum and renormalization prescription, the real and imaginary parts are fitted simultaneously over $z\in[10a,16a]=[0.76,1.22]~{\rm fm}$, using the NLA-inspired parametrization of Eq.~\eqref{eq:extrapolation_nla}. We allow $m\in[0.13,1.56]$~GeV and $n\in[0,4]$~\cite{Gao:2021dbh}, and take $z_{\rm extrap}=10a=0.76~{\rm fm}$. The directly determined lattice matrix elements are retained at $|z|<z_{\rm extrap}$, while the fitted asymptotic form is used at larger separations according to Eq.~\eqref{eq:piecewise_m}.

The resulting extrapolations are shown in Fig.~\ref{fig:renorm_extrap}. For the CG case, the matrix elements are already small at the largest measured separations, and the extrapolation should therefore not be interpreted as supplying a sizable unresolved physical tail. Its role is primarily to impose the expected smooth asymptotic falloff on residual long-distance fluctuations and thereby stabilize their contribution to the Fourier transform. We then Fourier transform the extended matrix elements according to Eq.~\eqref{eq:qv_qplus_reim}. The lattice-data region is integrated directly, while the fitted asymptotic form supplies the contribution over $z_{\rm extrap}\le |z|<\infty$. This yields the quasi-valence and full-quark distributions, $\widetilde q_v(x,P_z)$ and $\widetilde q_+(x,P_z)$, which form the input to the perturbative matching discussed in Sec.~\ref{subsec:matching}.

\begin{figure*}[!tp]
    \centering
    \includegraphics[width=0.47\linewidth]{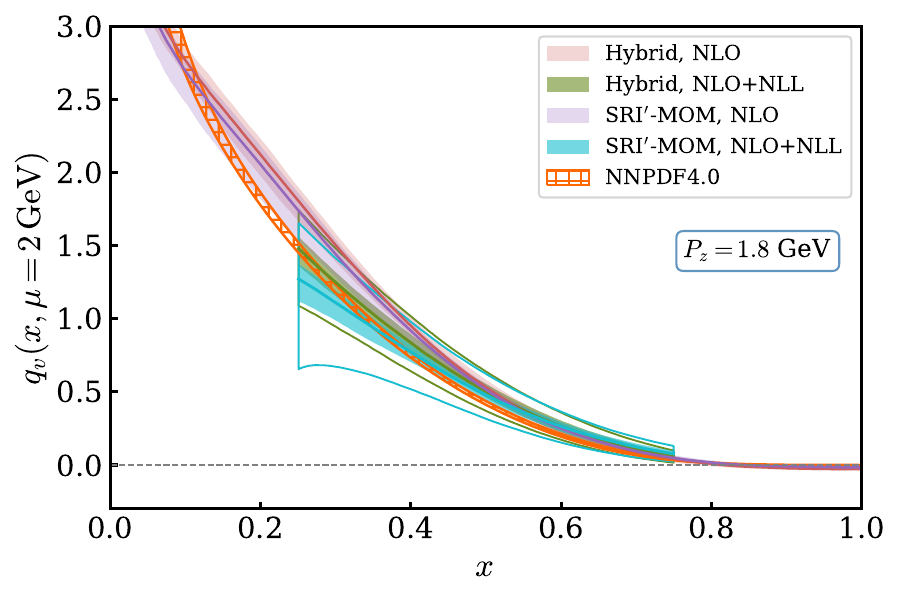}
    \includegraphics[width=0.47\linewidth]{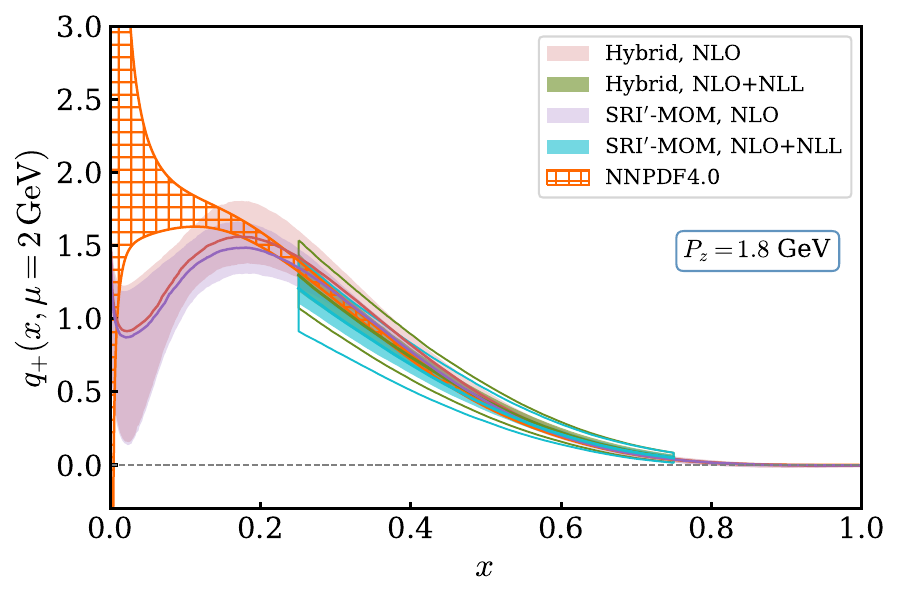} \\
    \includegraphics[width=0.47\linewidth]{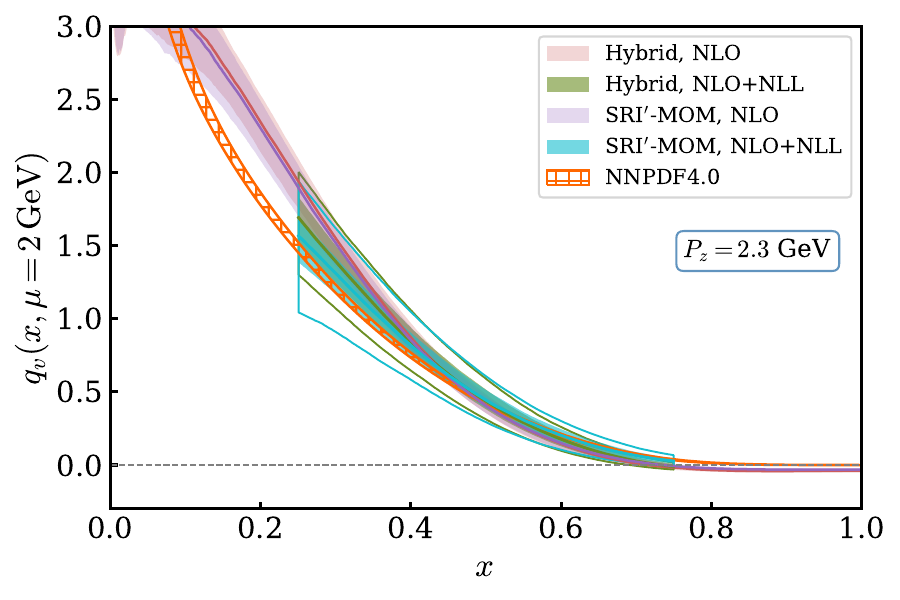}
    \includegraphics[width=0.47\linewidth]{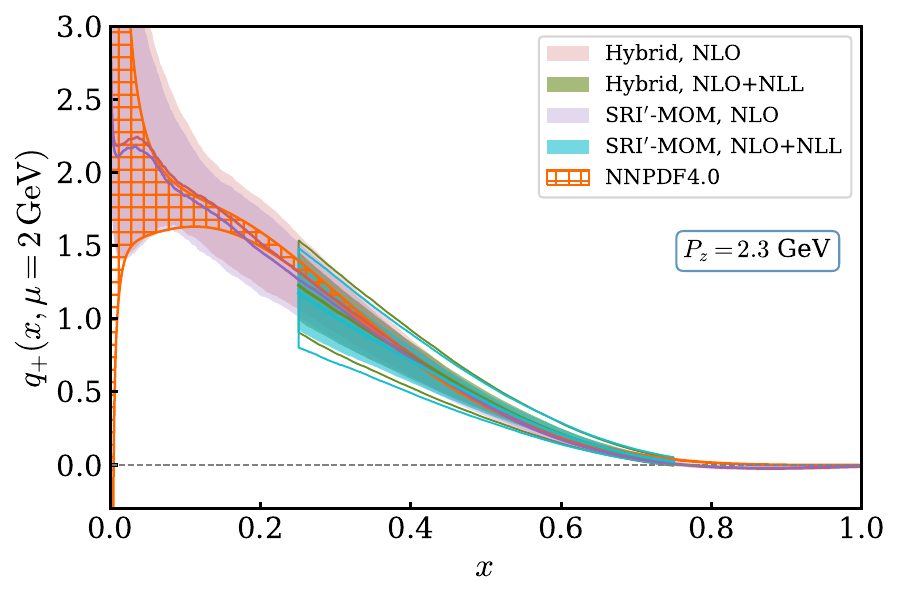}
    \caption{Matched isovector PDFs obtained from hybrid renormalization with $\zs=2a=0.152$~fm and from SRI$^\prime$-MOM renormalization, in the $\overline{\rm MS}$ scheme at $\mu=2$~GeV. Top: $P_z=1.8$~GeV; bottom: $P_z=2.3$~GeV. Left: $q_v$; right: $q_+$. Red and green show the hybrid NLO and NLO+NLL results, while purple and cyan show the corresponding SRI$^\prime$-MOM results. For NLO+NLL distributions, displayed over $0.25<x<0.75$, filled bands denote the propagated uncertainties for the central choice $\kappa=1$, and the outer contours include the variation $\kappa\in\{1/\sqrt{2},1,\sqrt{2}\}$. The orange cross-hatched band shows NNPDF4.0~\cite{NNPDF:2021njg}.}
    \label{fig:scheme_comparison}
\end{figure*}

\subsection{Light-Cone PDFs}
\label{subsec:pdf_results}

The quasi-PDFs obtained above are matched to the light-cone PDFs following Sec.~\ref{subsec:matching}. We first examine two internal consistency tests of the extraction: the dependence on the renormalization prescription and on the momentum. We then compare the highest-momentum results with earlier lattice calculations and with phenomenological global analyses.

Figure~\ref{fig:scheme_comparison} compares the isovector valence distribution $q_v$ and full-quark distribution $q_+$ obtained from SRI$^\prime$-MOM and hybrid renormalization at $P_z=1.8$ and $2.3$~GeV. Both fixed-order NLO and resummed NLO+NLL results are shown in the $\overline{\rm MS}$ scheme at $\mu=2$~GeV. The fixed-order results are shown alongside the resummed determinations to illustrate the numerical effect of the NLL evolution. The NNPDF4.0 distribution is included as a reference.

For quantitative comparisons, we restrict the resummed results to the common intermediate region $0.25<x<0.75$, where both the perturbative matching and the finite-momentum expansion are expected to be under better control, as discussed in Sec.~\ref{subsec:matching}. The difference between the NLO and NLO+NLL central values illustrates the numerical impact of the RGR treatment of the collinear logarithms associated with the partonic scale \(2|x|P_z\). Within the intermediate region, the SRI$^\prime$-MOM and hybrid NLO+NLL determinations are mutually compatible within the quoted uncertainty envelopes at both momenta. The dependence on the hybrid transition distance is examined explicitly in Appendix~\ref{app:hybrid_comparison}.

\begin{figure*}[!tp]
    \centering
    \includegraphics[width=0.47\linewidth]{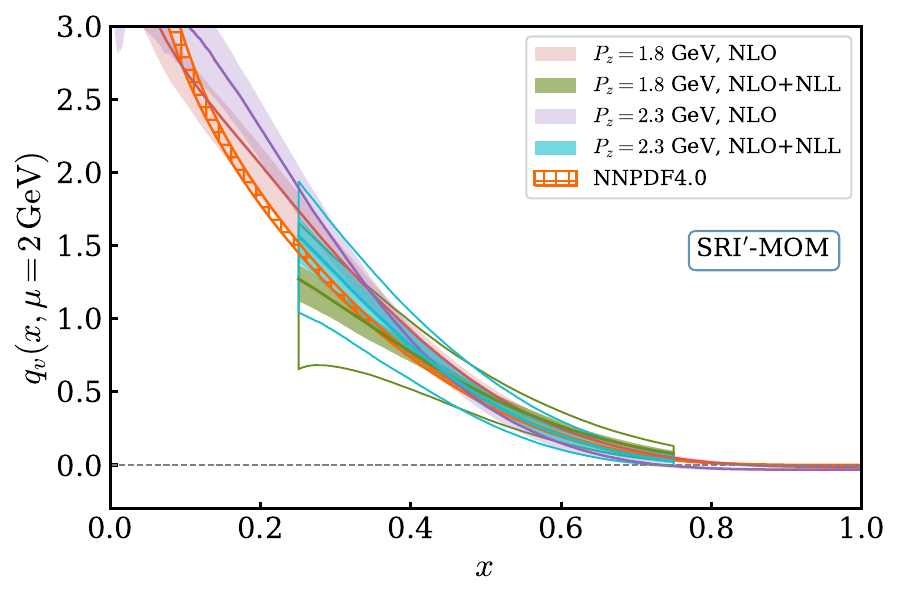}
    \includegraphics[width=0.47\linewidth]{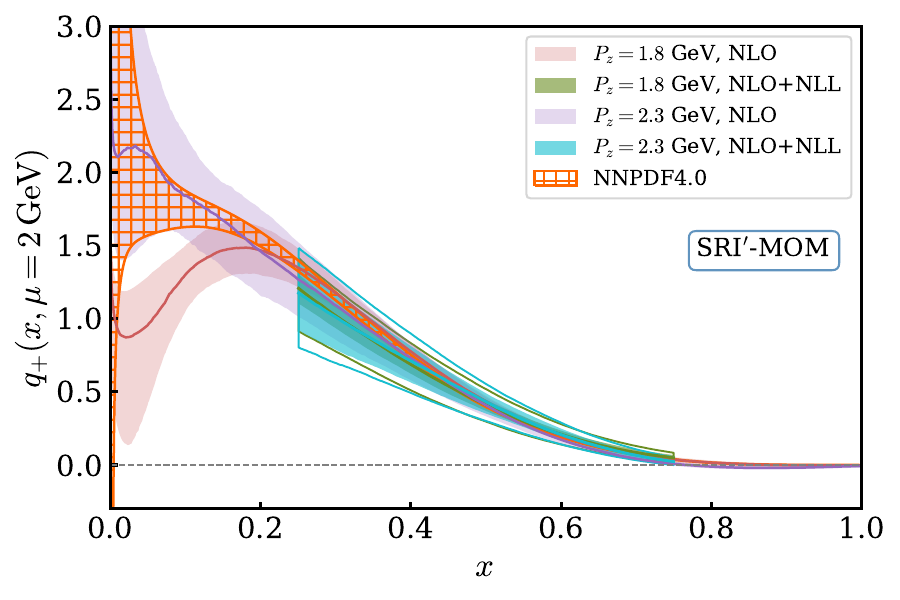} \\
    \includegraphics[width=0.47\linewidth]{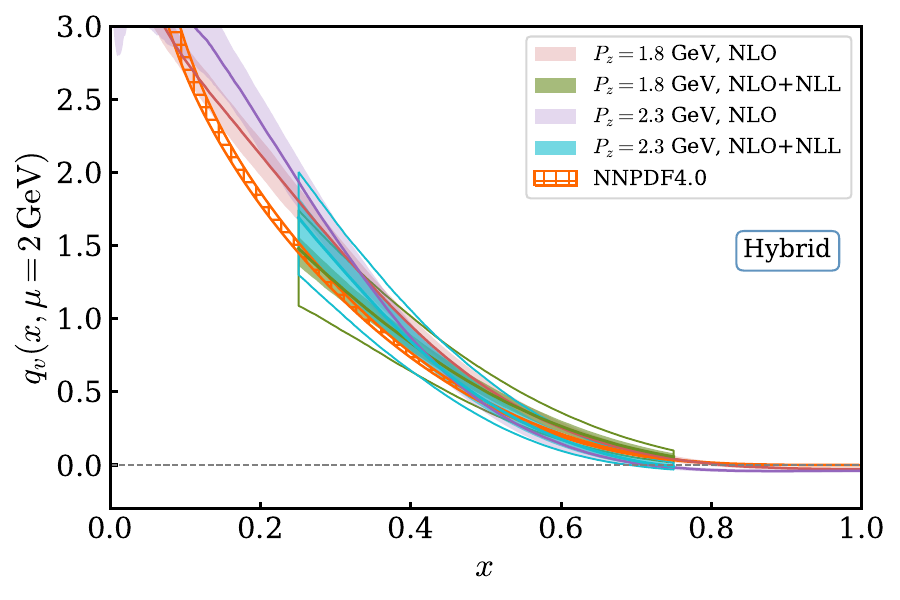}
    \includegraphics[width=0.47\linewidth]{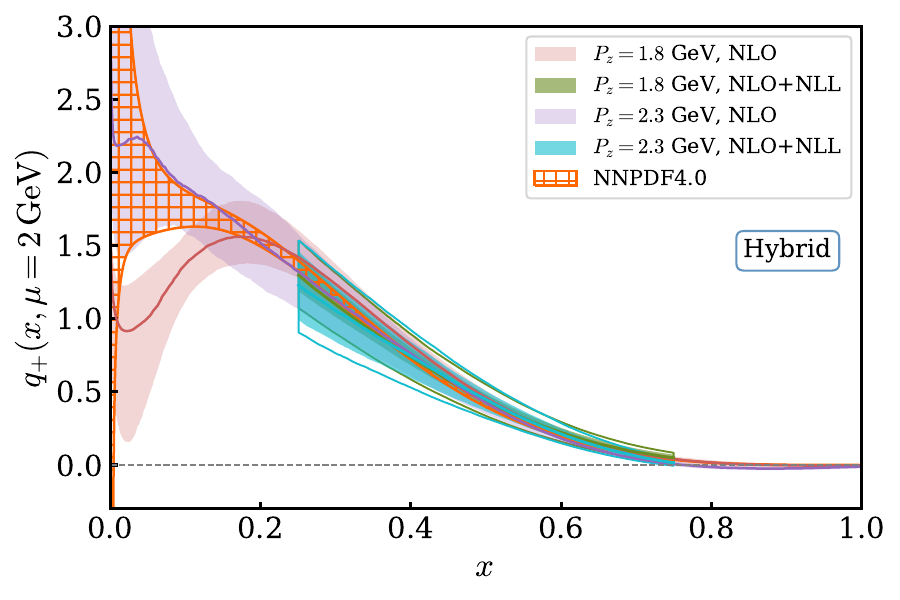}
    \caption{Momentum dependence of the isovector light-cone PDFs in the $\overline{\rm MS}$ scheme at $\mu=2$~GeV. Top: SRI$^\prime$-MOM; bottom: hybrid renormalization with $\zs=2a=0.152$~fm. Left: $q_v$; right: $q_+$. Red and green denote the NLO and NLO+NLL results at $P_z=1.8$~GeV, while purple and cyan denote the corresponding results at $P_z=2.3$~GeV. For NLO+NLL, filled bands show the propagated uncertainties at $\kappa=1$, and the outer contours include the variation $\kappa\in\{1/\sqrt{2},1,\sqrt{2}\}$. The orange cross-hatched band shows NNPDF4.0~\cite{NNPDF:2021njg}.}
    \label{fig:momentum_convergence}
\end{figure*}

We next examine the residual momentum dependence in Fig.~\ref{fig:momentum_convergence}. At fixed renormalization prescription, comparing the results at $P_z=1.8$ and $2.3$~GeV provides a direct test of the approach to the large-momentum regime after perturbative matching and evolution. Over the common intermediate-$x$ region, the NLO+NLL results at the two boosts are consistent within the displayed uncertainty envelopes for both renormalization prescriptions. This qualitative pattern is consistent with the expected reduction of finite-momentum corrections as the hadron boost is increased.

Having established the consistency between the two renormalization prescriptions and between the two available boosts, we combine the isovector valence and full-quark distributions according to Eq.~\eqref{eq:qv_qplus_lc}. For $x>0$,
\begin{equation}
    q_{u-d}(x)=u(x)-d(x), \quad q_{u-d}(-x)=\bar d(x)-\bar u(x).
\end{equation}
For the comparisons below, we use the highest available momentum, $P_z=2.3$~GeV.

\begin{figure*}[!tp]
    \centering
    \includegraphics[width=0.49\linewidth]{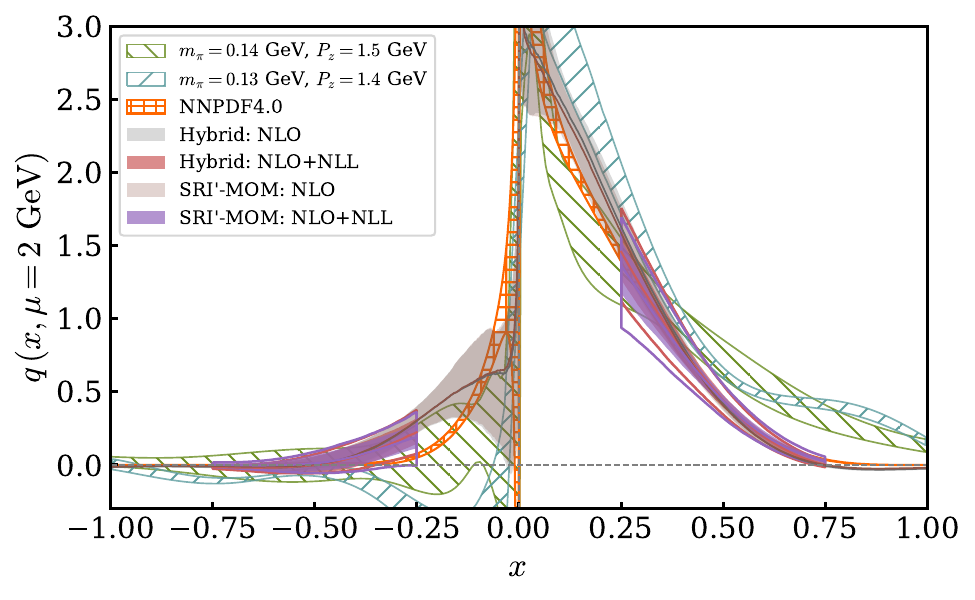}
    \hfill
    \includegraphics[width=0.49\linewidth]{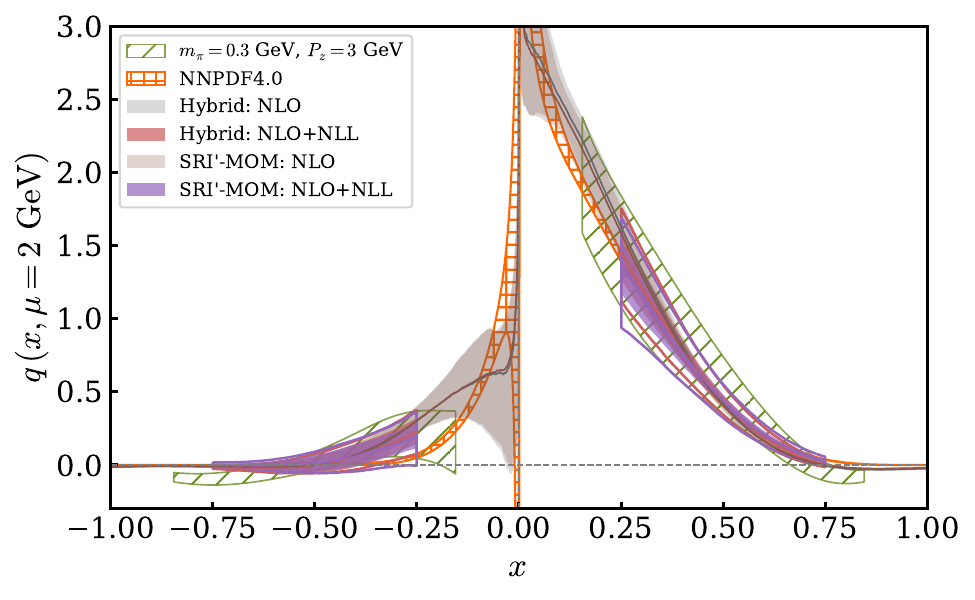}
    \caption{Comparison of the present isovector light-cone PDF with earlier lattice determinations, in the $\overline{\rm MS}$ scheme at $\mu=2$~GeV. Left: physical-quark-mass calculations at $P_z=1.4$~GeV~\cite{Alexandrou:2018pbm} and $P_z=1.5$~GeV~\cite{Gao:2022uhg}. Right: the CG calculation at $m_\pi=300$~MeV and $P_z=3$~GeV~\cite{Gao:2026hix}. In both panels, the present results are obtained at $P_z=2.3$~GeV for both renormalization schemes. Gray and red denote the hybrid NLO and NLO+NLL results, respectively, while brown and purple denote the corresponding SRI$^\prime$-MOM results. For NLO+NLL, filled bands show the propagated uncertainties at $\kappa=1$, and outer contours include matching-scale variation. The orange cross-hatched band shows NNPDF4.0~\cite{NNPDF:2021njg}.}
    \label{fig:lattice_comparison}
\end{figure*}

Figure~\ref{fig:lattice_comparison} compares the present determinations with representative previous lattice calculations. The left panel shows physical-quark-mass calculations performed at lower momenta, $P_z=1.4$ and $1.5$~GeV, while the right panel shows the recent CG calculation at $m_\pi=300$~MeV and $P_z=3$~GeV. The earlier physical-mass calculations at lower boosts exhibit visible differences from NNPDF4.0 over parts of the momentum-fraction range, whereas our physical-mass large-momentum results and the heavier-mass large-momentum calculations are broadly compatible with the phenomenological distribution over the intermediate-$|x|$ region within their respective uncertainties. Taken together, this pattern suggests that reaching sufficiently large momentum plays a more prominent role in the observed agreement with phenomenology than the pion-mass difference between the physical one and \(m_\pi=300\) MeV. Because the lattice setups and analysis procedures also differ, this comparison supports that interpretation only qualitatively.

\begin{figure}[!t]
    \centering
    \includegraphics[width=\columnwidth]{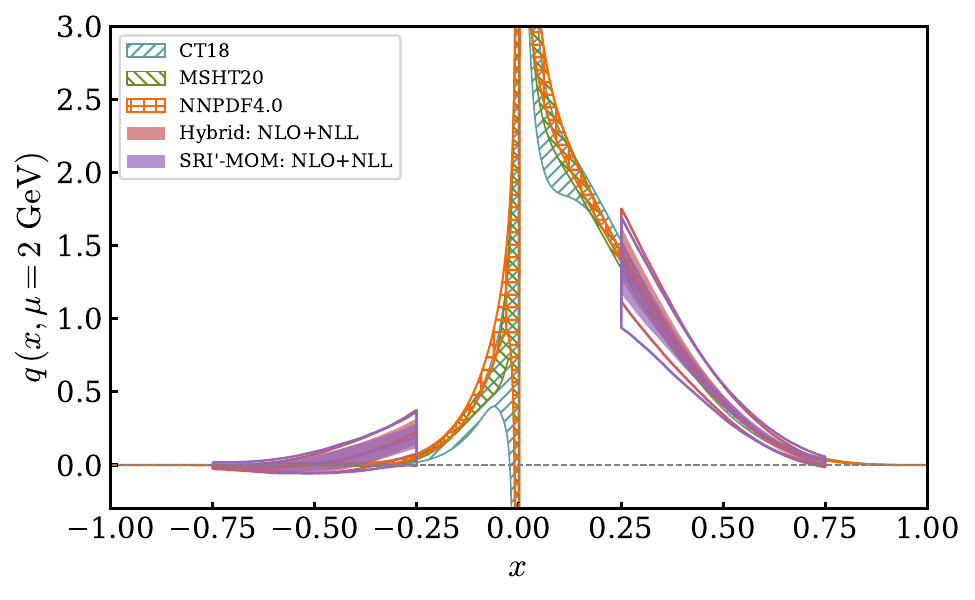}
    \caption{Isovector light-cone PDFs obtained from the SRI$^\prime$-MOM and hybrid ($\zs=2a$) analyses at $P_z=2.3$~GeV, compared with CT18, MSHT20, and NNPDF4.0~\cite{Hou:2019efy,Bailey:2020ooq,NNPDF:2021njg}. All distributions are shown in the $\overline{\rm MS}$ scheme at $\mu=2$~GeV. For $x>0$, $q(x)=u(x)-d(x)$, while $q(-x)=\bar d(x)-\bar u(x)$. Red and purple denote the NLO+NLL results obtained with the hybrid and SRI$^\prime$-MOM prescriptions, respectively. Filled bands show the propagated uncertainties at $\kappa=1$, and the outer contours include matching-scale variation.}
    \label{fig:entire_x}
\end{figure}

We finally compare our highest-momentum determination with the global analyses CT18, MSHT20, and NNPDF4.0 in Fig.~\ref{fig:entire_x}. The NLO+NLL lattice PDFs from both renormalization prescriptions are compatible with all three phenomenological determinations over the studied intermediate-$|x|$ region, within the matching-scale variation. In the antiquark region, the lattice central values lie somewhat above the global-fit central values over part of this window. Since $q_{u-d}(-x)=\bar d(x)-\bar u(x)$ for $x>0$, the negative-$x$ result favors an excess of $\bar d$ over $\bar u$ and is qualitatively consistent with the experimentally observed flavor asymmetry of the proton sea~\cite{NewMuon:1991hlj,NuSea:2001idv,SeaQuest:2021zxb}. Taken together, the agreement with phenomenology in coordinate and momentum space provides compelling evidence for the reliability of the CG approach to lattice calculations of the proton isovector unpolarized PDF.

\section{\label{sec:summary}Summary}
We have calculated the proton isovector ($u-d$) unpolarized PDF at physical quark masses within the LaMET CG framework, using a lattice spacing of $a=0.076$~fm and proton momenta $P_z=1.8$ and $2.3$~GeV. Kinematically enhanced interpolating operators improve the signal quality at large boosts, while CG quark bilinears avoid the Wilson-line linear divergence and its associated renormalon ambiguity. We combine two complementary ground-state extractions to incorporate the associated method dependence. The matrix elements are renormalized using SRI$^\prime$-MOM followed by conversion to $\overline{\rm MS}$, and, separately, using the hybrid prescription. For a complementary coordinate-space comparison, we reconstruct matrix elements from NNPDF4.0 using the fixed-order NLO matching kernel appropriate to each renormalization prescription, followed by an inverse Fourier transform. After long-distance continuation, we Fourier transform the renormalized matrix elements to obtain quasi-PDFs and match them to light-cone PDFs in the $\overline{\rm MS}$ scheme at $\mu=2$~GeV using NLO matching, both at fixed order and supplemented by NLL RGR. We then compare the resulting light-cone PDFs with phenomenological global analyses in momentum space.

Our principal result is the simultaneous agreement with phenomenology in both coordinate and momentum space. The NLO+NLL PDFs are compatible with phenomenological determinations over $0.25\lesssim|x|\lesssim0.75$, extending to moderately large $|x|$, within the displayed uncertainty envelopes. The coordinate-space matrix elements also agree with NNPDF4.0-based NLO reconstructions at short quark separations, with particularly close agreement in the imaginary part observed for the first time. The compatibility of the PDF results at the two momenta and with the two renormalization prescriptions provides additional checks within the present analysis. Additionally, the isovector antiquark sector favors a positive $\bar{d}-\bar{u}$ signal consistent with the flavor-asymmetric nucleon sea inferred from the NMC, NuSea, and SeaQuest experiments~\cite{NewMuon:1991hlj, NuSea:2001idv, SeaQuest:2021zxb}.

Together with our previous calculation~\cite{Gao:2026hix}, these encouraging results demonstrate the reliability of the CG approach and mark another firm step toward precision benchmarking of lattice calculations of PDFs. We also note that the current perturbative accuracy of the CG method is at NLO, which is expected to be improved to next-to-next-to-leading order in the near future, allowing us to estimate the effects of higher-order corrections. Besides, threshold resummation will improve the large logarithms at $|x|\to1$. The residual $z_s$-dependence in the hybrid scheme motivates higher-order calculations of the short-distance coefficient $C_0$ and the associated matching kernel, together with an investigation of renormalon effects and their subtraction where appropriate. Meanwhile, SRI$^\prime$-MOM has been demonstrated to be a reliable renormalization scheme within the CG approach and is worth further development. Comparisons at finer lattice spacings would further help assess discretization effects. Further studies of excited-state effects, together with calculations at additional lattice spacings and volumes, will enable systematic improvements such as continuum and infinite-volume extrapolations and enable more stringent comparisons with phenomenology.

\acknowledgments
We thank Fei Yao and Rui Zhang for valuable discussions. We also thank Krzysztof Cichy for providing the numerical results from Ref.~\cite{Alexandrou:2018pbm}.

This material is based upon work supported by the U.S.~Department of Energy, Office of Science, Office of Nuclear Physics, through Contract No.~DE-SC0012704, Contract No.~DE-AC02-06CH11357, within the framework of the Scientific Discovery through Advanced Computing (SciDAC) award Fundamental Nuclear Physics at the Exascale and Beyond, and the Quark-Gluon Tomography (QGT) Topical Collaboration with Award DE-SC0023646. YZ also acknowledges support from the U.S. Department of Energy, Office of Science, Office of Nuclear Physics, through the Early Career Award under Contract No.~DE-SCL0000017. QS is supported by Kent State University through index 201452.
This manuscript has been authored by Fermi Forward Discovery Group, LLC under Contract No. 89243024CSC000002 with the U.S. Department of Energy, Office of Science, Office of High Energy Physics.

This research used awards of computer time provided by the INCITE program at the Argonne Leadership Computing Facility, a DOE Office of Science User Facility operated under Contract DE-AC02-06CH11357; the ALCC program at the Oak Ridge Leadership Computing Facility, which is a DOE Office of Science User Facility supported under Contract DE-AC05-00OR22725; and the National Energy Research Scientific Computing Center, a DOE Office of Science User Facility supported by the Office of Science of the U.S.~Department of Energy under Contract DE-AC02-05CH11231, using NERSC awards NP-ERCAP0032114, NP-ERCAP0035891 and NP-ERCAP0036132.

The gauge ensemble used in this work was generated with the \texttt{SIMULATeQCD} software suite~\cite{HotQCD:2023ghu}. Coulomb-gauge fixing was performed with \href{https://github.com/RJHudspith/GLU}{\texttt{GLU}}~\cite{Hudspith:2014oja}. The quark propagators and correlation functions were computed using \texttt{PyQUDA}~\cite{Jiang:2024lto}, which interfaces with the \texttt{QUDA} library and employs its multigrid solver~\cite{Clark:2009wm,Babich:2011np,Clark:2016rdz}.

\appendix
\section{\label{app:me_extraction}Ground-state fits}

\begin{figure}[!tp]
    \centering
    \includegraphics[width=0.86\linewidth]{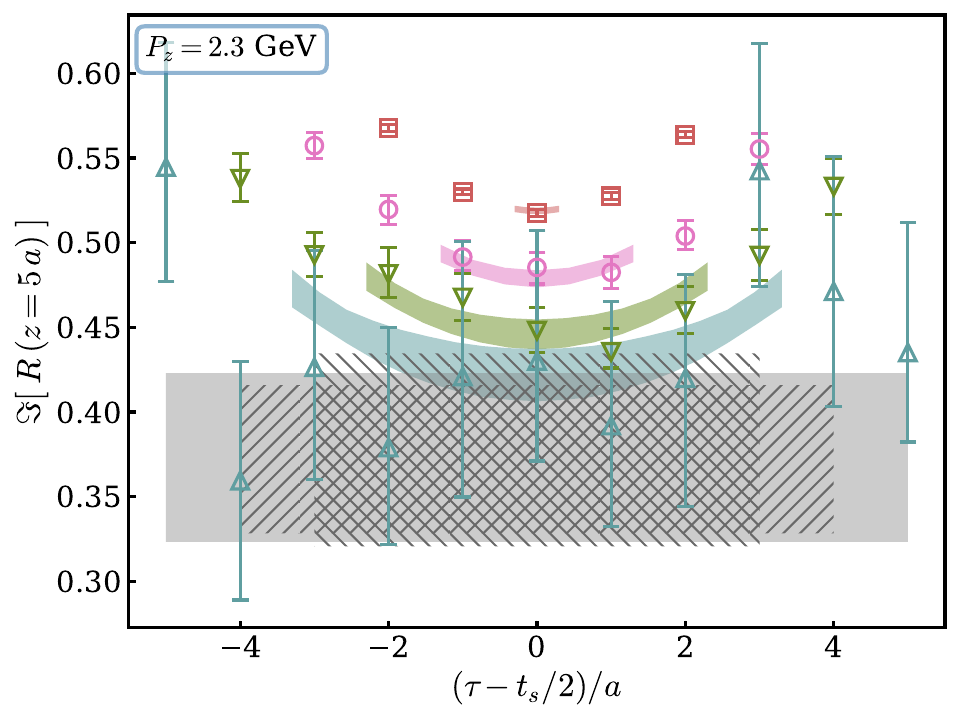}\\[-0.5ex]
    \includegraphics[width=0.86\linewidth]{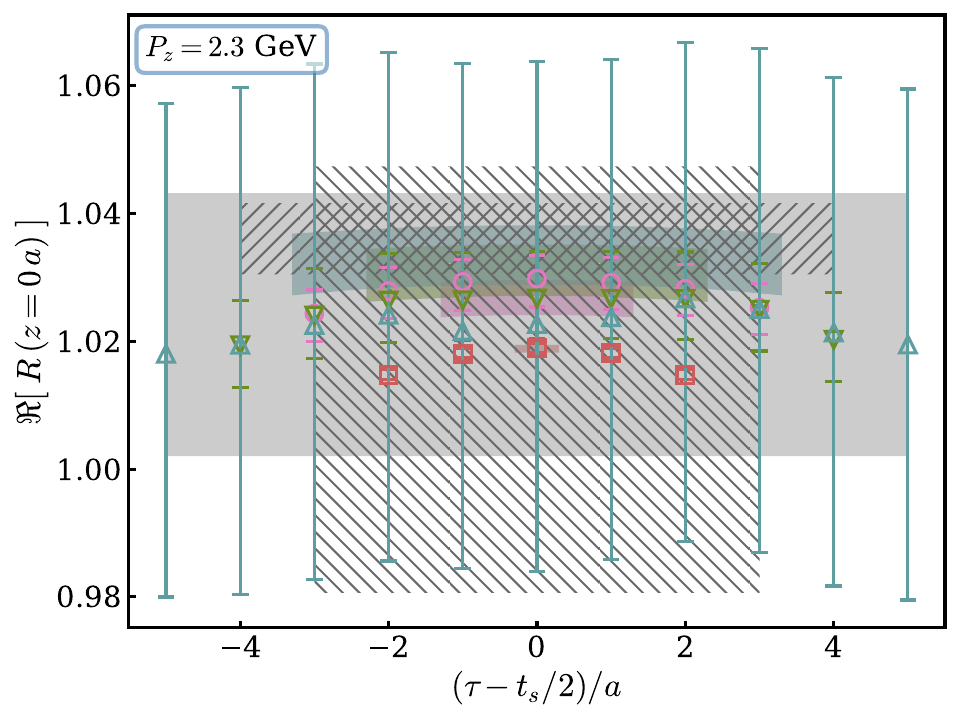}\\[-0.5ex]
    \includegraphics[width=0.86\linewidth]{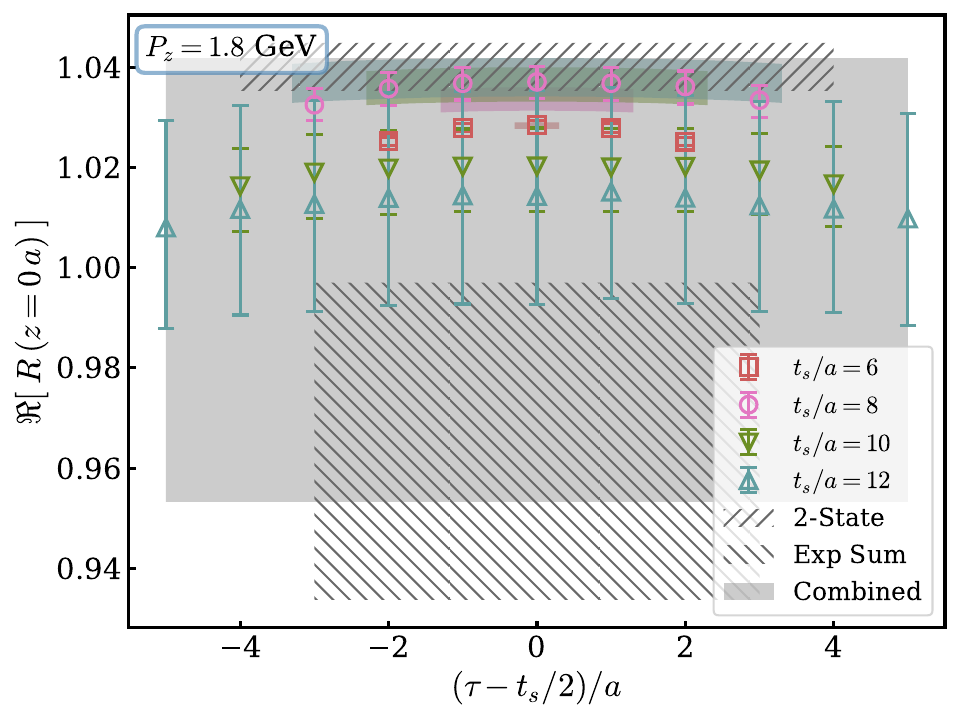}
    \caption{\label{fig:ratio_sample} Isovector lattice ratios $R(\mathbf{P};\tau,\ts;z)$ and the extracted ground-state bare matrix elements for three representative cases. Open symbols show the lattice data at the four source-sink separations $\ts/a \in \{6,8,10,12\}$, and the colored bands show the corresponding ratios reconstructed from the two-state fit. The two hatched bands denote the ground-state matrix elements obtained from the two-state and exponential summation fits, while the solid gray band labeled ``Combined'' shows the central $68\%$ interval of their equally weighted combined bootstrap distribution. From top to bottom: the imaginary part at $P_z=2.3$~GeV and $z=5a$, the real part at $P_z=2.3$~GeV and $z=0$, and the real part at $P_z=1.8$~GeV and $z=0$. The panels are ordered from better to poorer agreement between the extraction methods; the kinematic labels identify the examples but do not imply a systematic dependence on momentum or separation.}
\end{figure}

The asymptotic limit of the ratio in Eq.~\eqref{eq:ratio} cannot be reached directly because the SNR deteriorates exponentially with increasing $\ts$. At the finite separations accessible in the calculation, excited states can contribute when the operator insertion approaches either the source or the sink. We therefore analyze the three-point functions at $\ts/a\in\{6,8,10,12\}$ and omit $\tau_{\rm skip}/a=3$ time slices adjacent to both boundaries. Two complementary methods are applied to the same underlying data and bootstrap samples:

\begin{enumerate}
    \item \textit{Two-state fit.} Truncating the spectral decomposition to the ground and first excited states, $m,n\in\{0,1\}$, gives
    \begin{equation}
    \begin{aligned}
        R(\tau,\ts)&\approx\frac{H(\tau,\ts)}{1+r^2 e^{-\Delta E\ts}},\\
        H(\tau,\ts)&=h_{00}+h_{11}r^2e^{-\Delta E\ts}\\
        &\quad+h_{01}r e^{-\Delta E\tau}
              +h_{10}r e^{-\Delta E(\ts-\tau)}.
    \end{aligned}
    \end{equation}
    Here $r=|A_1|/|A_0|$ and $\Delta E=E_1-E_0$. The energies and overlap amplitudes obtained from the two-point-function analysis are used as inputs to a joint fit over the available $\ts$ and $\tau\in[\tau_{\rm skip},\ts-\tau_{\rm skip}]$, from which we determine $h_{00}=h^{\rm B}$.

    \item \textit{Exponential summation fit.} We define the summed ratio
    \begin{equation}
        R_{\rm sum}(\ts)=\sum_{\tau=\tau_{\rm skip}}^{\ts-\tau_{\rm skip}}R(\tau,\ts),
    \end{equation}
    which suppresses the $\tau$-dependent transition contributions. Following the exponential summation ansatz of Ref.~\cite{Gao:2020ito}, we fit
    \begin{equation}
        R_{\rm sum}(\ts)\approx n h_{00}+B_1e^{-\Delta E\ts}+B_0,
    \end{equation}
    where $n=(\ts-2\tau_{\rm skip})/a+1$ is the number of time slices in the sum and $B_0$ and $B_1$ are independent fit parameters. The coefficient of $n$ determines $h_{00}=h^{\rm B}$.
\end{enumerate}

To mitigate autocorrelations, measurements on the 700 gauge configurations are rebinned in groups of 10, yielding 70 bins, from which we generate 400 bootstrap samples. Both extraction methods are evaluated on the same bootstrap ensembles. Their samples are then concatenated with equal weight to form the combined distribution used in the main analysis. These combined samples are propagated through renormalization, the long-distance continuation, the Fourier transform, and perturbative matching.

Figure~\ref{fig:ratio_sample} compares the two extraction strategies in three cases selected to span representative levels of agreement, rather than to diagnose a systematic dependence on $P_z$ or $z$. From top to bottom, the two ground-state estimates evolve from close agreement, through a modest method-dependent shift, to a visible separation; in the last case, the combined interval broadens to reflect the separation between the two determinations. These examples illustrate how the bootstrap aggregation retains the extraction-model dependence as the two fits become less consistent.

\section{\label{app:srimom}Determination of the SRI$^\prime$-MOM renormalization factor}

\begin{figure}[!htbp]
    \centering
    \includegraphics[width=0.94\linewidth]{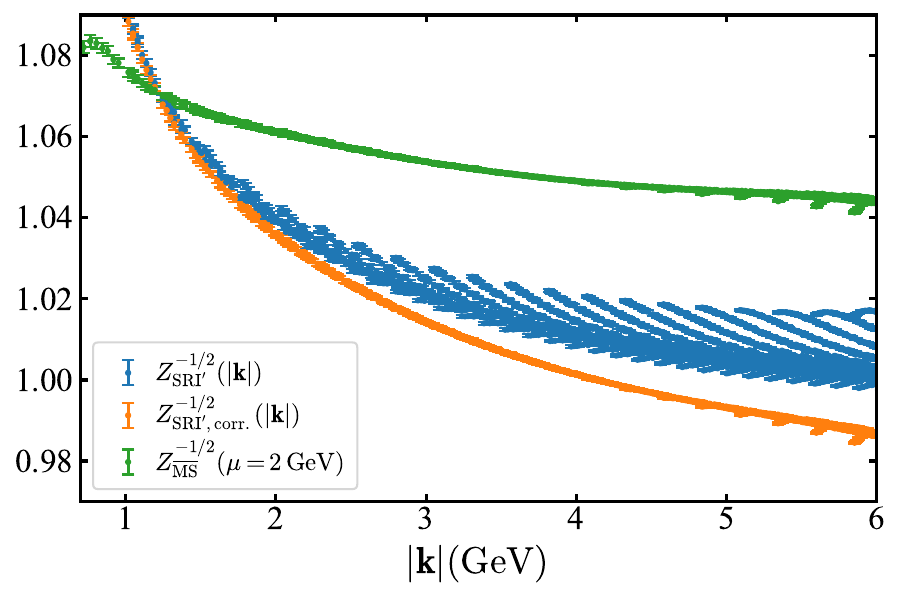}
    \caption{Determination of the CG quark-field renormalization factor. Blue points show $Z_{\rm SRI^\prime}^{-1/2}(|\mathbf{k}|)$, orange points show the result after removal of the leading cubic-symmetry-breaking artifacts, and green points show $Z_{\overline{\rm MS}}^{-1/2}(\mu=2~{\rm GeV})$ after running and conversion.}
    \label{fig:zpsi_msbar}
\end{figure}

Following Ref.~\cite{Gao:2026hix}, we implement the SRI$^\prime$-MOM prescription on the present ensemble. The mass-independent renormalization condition and perturbative conversion are defined in the chiral limit, while we evaluate the nonperturbative factor at the simulated quark mass. We remove the leading cubic-symmetry-breaking artifacts, evolve the corrected factors at leading-logarithmic (LL) accuracy, and apply the finite one-loop conversion to $\overline{\rm MS}$~\cite{Popovici:2008ty}. Figure~\ref{fig:zpsi_msbar} shows the renormalization factors before and after artifact removal and after conversion to the $\overline{\rm MS}$ scheme at $\mu=2$~GeV, plotted as $Z^{-1/2}$. The quark subtraction momentum $|\mathbf{k}|$ is distinct from the proton boost $P_z$. We observe that the $|\mathbf{k}|$ dependence of the SRI$^\prime$-MOM factor appears stronger than the logarithmic dependence predicted by the LL perturbation theory. For this reason, the renormalization factor in the $\overline{\rm MS}$ scheme still depends on the momentum $|\mathbf{k}|$. If the perturbative matching fully captured the $|\mathbf{k}|$ dependence, no such dependence should appear. 

\begin{figure*}[!tp]
    \centering
    \includegraphics[width=0.98\textwidth]
        {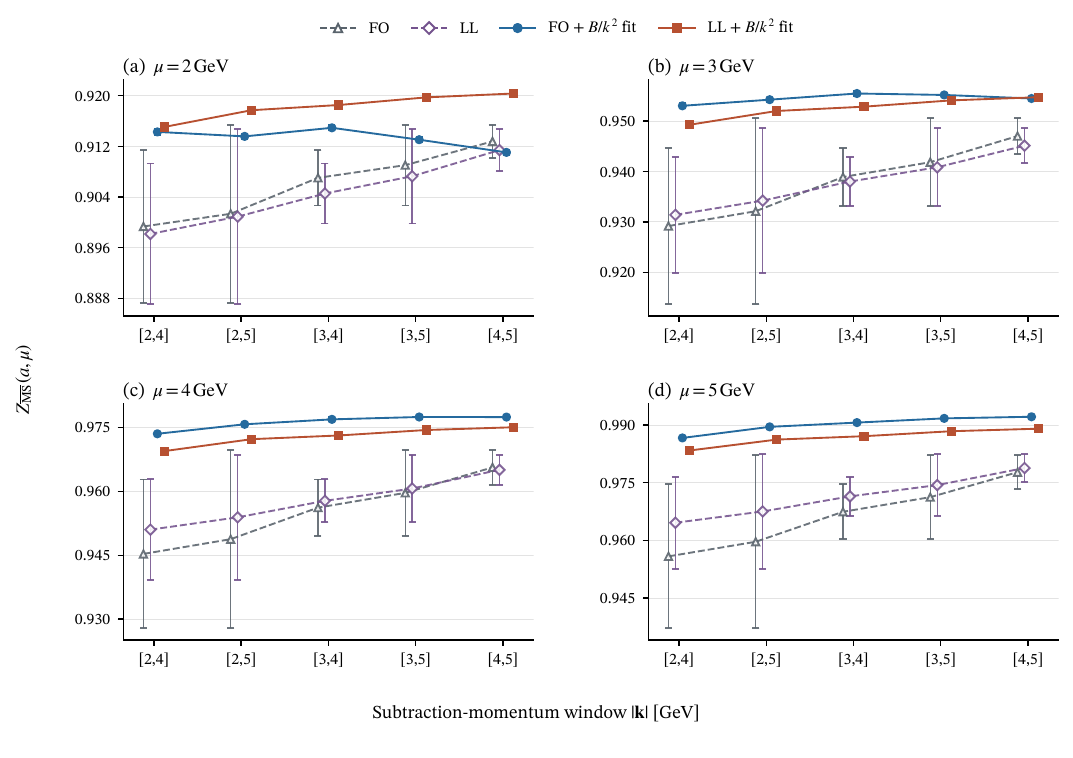}
    \caption{Dependence of $Z_{\overline{\rm MS}}(a,\mu)$ on the subtraction-momentum window at $\mu=2$, $3$, $4$, and $5$~GeV. FO denotes fixed-order one-loop conversion; LL denotes leading-logarithmic evolution followed by finite one-loop conversion. In the fitted analyses, a $B/|\mathbf{k}|^2$ term is added to the corrected SRI$^\prime$-MOM $Z^{-1/2}$ data.}
    \label{fig:zpsi_method_comparison}
\end{figure*}

Notably, in both our previous calculation on the $a=0.06$~fm lattice ensemble~\cite{Gao:2026hix} and an independent calculation~\cite{Grebe:2026qmt} using a MILC ensemble~\cite{MILC:2012znn} at $a=0.06$~fm, the $|\mathbf{k}|$ dependence almost completely cancels between the SRI$^\prime$-MOM factor and the LL matching factor. The remnant $|\mathbf{k}|$ dependence on the current lattice ensemble may be explained by the contribution from the dimension-two gluon condensate in fixed-gauge calculations of the form $B/|\mathbf{k}|^2$~\cite{Chetyrkin:2009kh,Blossier:2010ky}. To determine the renormalization factor in the $\overline{\rm MS}$ scheme, we evaluate the converted factor in several $|\mathbf{k}|$ bins. Figure~\ref{fig:zpsi_method_comparison} compares fixed-order (FO) and LL conversions, each with and without an empirical $B/|\mathbf{k}|^2$ term fitted to the corrected SRI$^\prime$-MOM data. For each ``no-fit'' determination, we take the midpoint of the envelope of the pointwise one-standard-deviation intervals in $Z_{\overline{\rm MS}}$ as the central value, with the corresponding half-width shown as its error bar. The fitted error bars are bootstrap statistical uncertainties conditional on the $B/|\mathbf{k}|^2$ ansatz. One can see that including the $B/|\mathbf{k}|^2$ term leads to $\overline{\rm MS}$ renormalization factors with very mild dependence on the subtraction-momentum window. As the windows shift to larger $|\mathbf{k}|^2$, the difference between determinations with and without the $B/|\mathbf{k}|^2$ term decreases, as expected.

For our final result at $\mu=2$~GeV, we choose $4\leq|\mathbf{k}|\leq5$~GeV as a compromise between reducing sensitivity to low-momentum effects and limiting potential finite-$a$ effects at larger $a|\mathbf{k}|$. The LL no-fit result in this window gives $Z_{\overline{\rm MS}}(a,2~{\rm GeV})=0.9114$. The LL fit including $B/|\mathbf{k}|^2$ in the same window gives $0.9204$. We assign their absolute difference, $0.0089$, as a symmetric uncertainty, obtaining $Z_{\overline{\rm MS}}(a,2~{\rm GeV})=0.9114(89)$. This comparison quantifies sensitivity to the analysis prescription; the fitted term does not establish a physical power correction, since higher-order discretization effects and omitted perturbative terms may also contribute to the observed momentum dependence. We have also varied the ${\overline{\rm MS}}$ scale $\mu$ from 2~GeV to 5~GeV and repeated the analysis, and the conclusions remain the same. At $\mu=2$~GeV, the difference between FO and LL fits with the $B/|\mathbf{k}|^2$ term is the largest within the window $[4,5]$, but it is still within the estimated uncertainty of the ${\overline{\rm MS}}$ renormalization factor.

\section{\label{app:hybrid_comparison}Transition-distance dependence of hybrid renormalization in Coulomb gauge}

\begin{figure*}[!tp]
    \centering
    \includegraphics[width=0.47\linewidth]{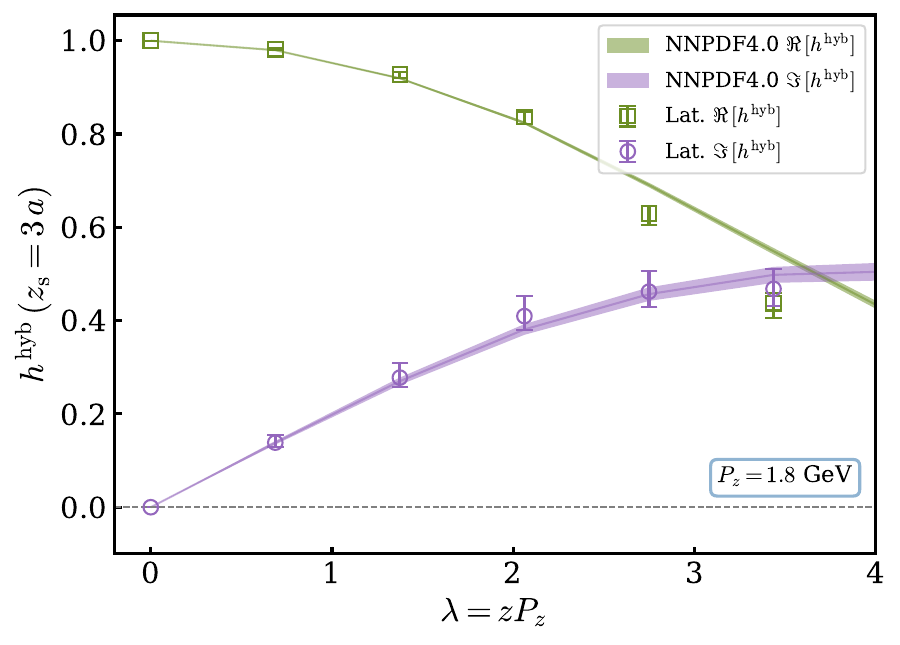}
    \includegraphics[width=0.47\linewidth]{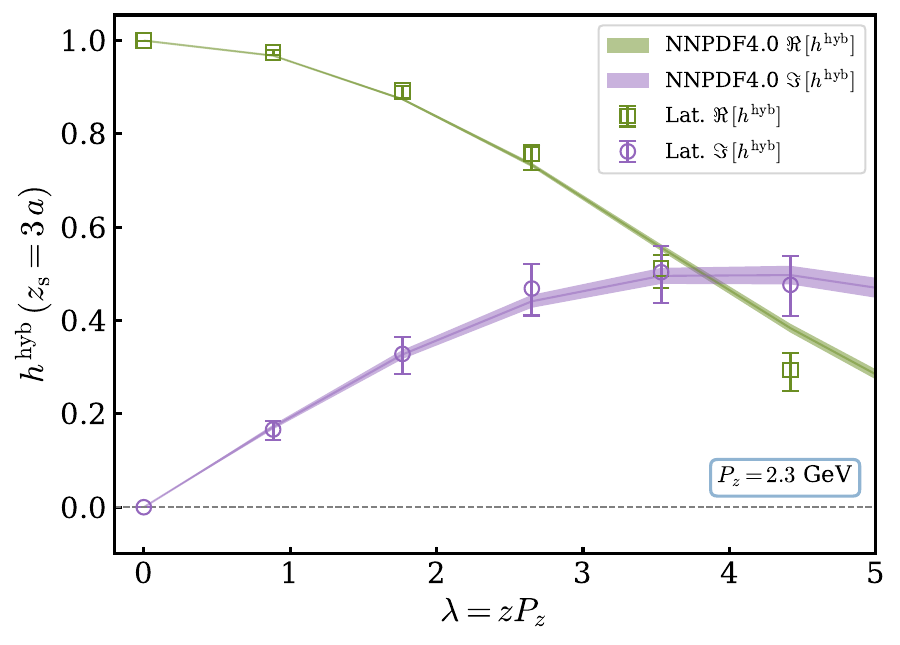} \\
    \includegraphics[width=0.47\linewidth]{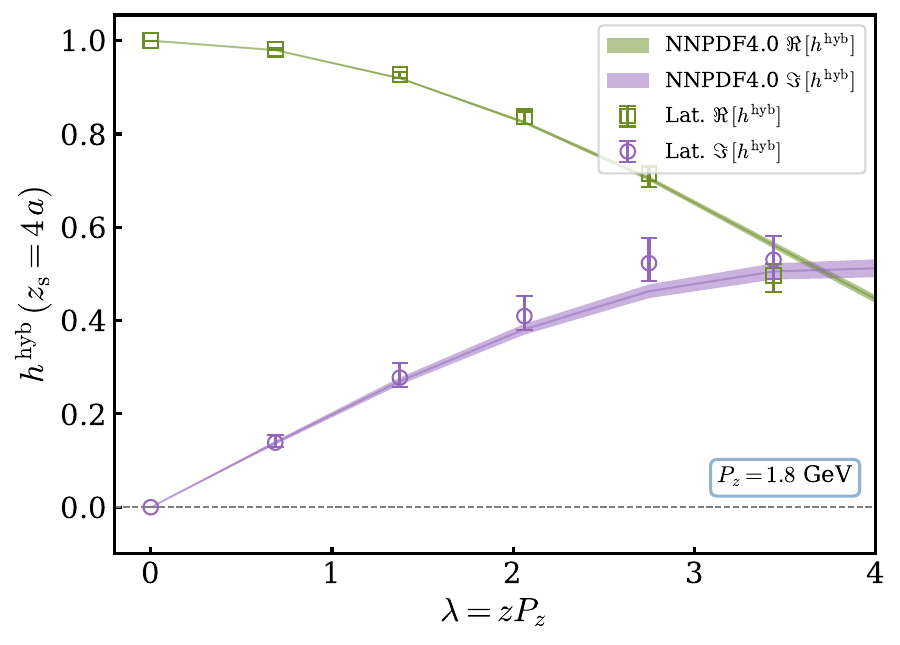}
    \includegraphics[width=0.47\linewidth]{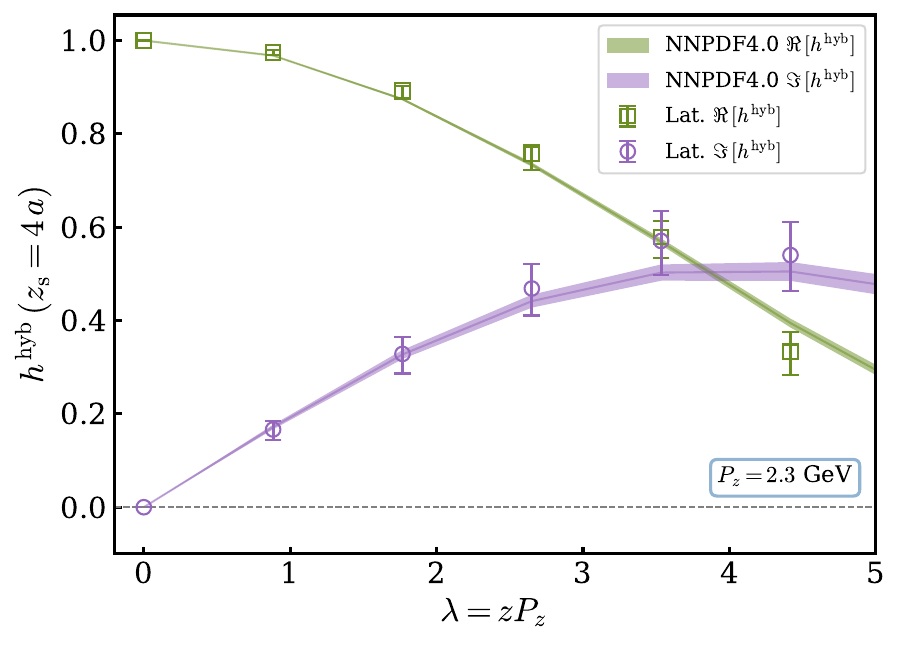}
    \caption{Hybrid-renormalized coordinate-space matrix elements compared with NNPDF4.0-based reconstructions using NLO CG matching. Top: $\zs=3a$; bottom: $\zs=4a$. Left: $P_z=1.8$~GeV; right: $P_z=2.3$~GeV. Green squares and purple circles denote the real and imaginary lattice matrix elements, respectively. The corresponding bands show the reconstructions from NNPDF4.0.}
\label{fig:hybrid_zs_coordinate}
\end{figure*}

The transition distance $\zs$ is an auxiliary parameter of the hybrid prescription. In the continuum and at leading power, its dependence is expected to cancel between the hybrid-renormalized quasi-PDF and the corresponding matching kernel when both are treated consistently to sufficient perturbative accuracy~\cite{Ji:2020brr, Gao:2023lny}. The absence of a spatial Wilson line in the CG operator removes the Wilson-line linear divergence, but it does not remove short-distance power corrections, finite-lattice-spacing effects, or perturbative truncation errors. We examine the resulting residual dependence by comparing $\zs=2a$, $3a$, and $4a$ at $P_z=1.8$ and $P_z=2.3$~GeV, both in coordinate and momentum spaces. Figure~\ref{fig:hybrid_zs_coordinate} compares the hybrid-renormalized matrix elements for $\zs=3a$ and $4a$ with NNPDF4.0-based reconstructions using the corresponding hybrid matching kernels. At both momenta, agreement is retained at short distances. For $|z|\leq2a$, all three transition distances use the same short-distance branch of Eq.~\eqref{eq:hybrid_cg}; the comparisons in this common region therefore share the same lattice matrix elements.

The zero-momentum reference matrix element is central to the $\zs$ comparison. For $|z|>\zs$, its value at $\zs$ fixes the normalization of the entire long-distance branch in Eq.~\eqref{eq:hybrid_cg}. Its relative separation dependence can be tested against the short-distance coefficient \(C_0(z,\mu)\equiv C_0^{{\rm CG},\gamma^t}(z,\mu)\)~\cite{Gao:2023lny,Gao:2026hix}:
\begin{equation}
    C_0(z,\mu)=1+\frac{\alphas(\mu)C_F}{4\pi}
      \left[1-\ln\!\frac{z^2\mu^2e^{2\gamma_E}}{4}\right]
      +\mathcal{O}(\alphas^2).
    \label{eq:c0_cg}
\end{equation}
Ratios at adjacent nonzero separations are renormalization group invariant. For $R_{\rm lat}(z)\equiv h^{\rm B}(z,0)/h^{\rm B}(z-a,0)$, the perturbative counterpart at leading power in the continuum is
\begin{equation}
    R_{\rm lat}(z) \simeq\frac{C_0(z,\mu)}{C_0(z-a,\mu)} = 1-\frac{\alphas(\mu)C_F}{2\pi}
       \ln\!\frac{z}{z-a}+\mathcal{O}(\alphas^2).
    \label{eq:c0_adjacent_ratio}
\end{equation}

\begin{figure}[!htbp]
    \centering
    \includegraphics[width=0.94\linewidth]{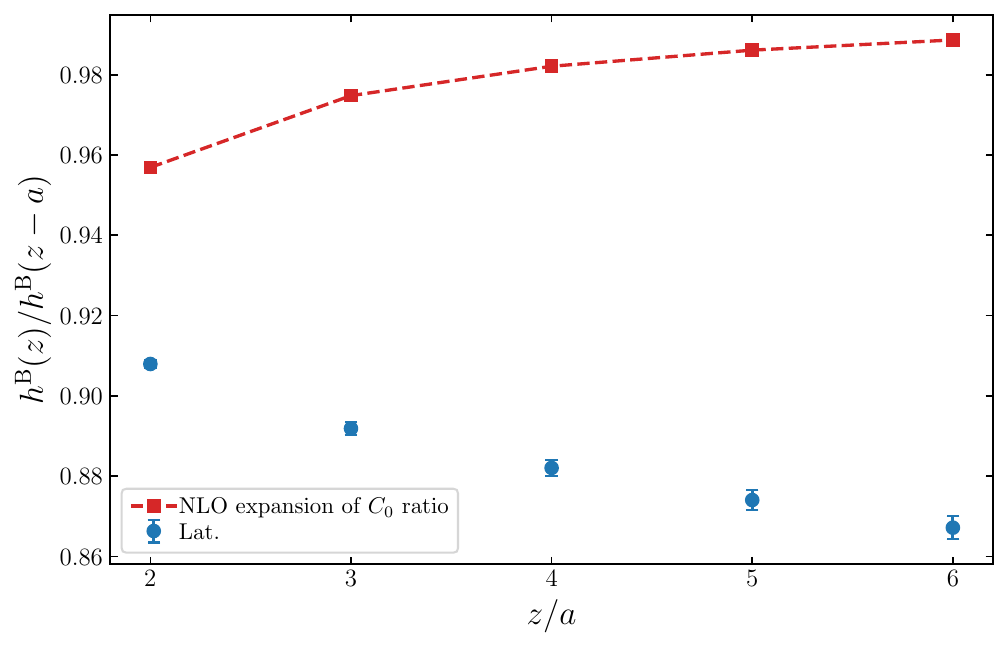}
    \caption{Adjacent ratios of the zero-momentum isovector bare matrix element on the $a=0.076$~fm ensemble (blue circles), compared with the strict NLO expansion of $C_0(z,\mu)/C_0(z-a,\mu)$ (red squares and dashed line). The strict-NLO curve is evaluated using \(\alphas(\mu=2~{\rm GeV})=0.293\).}
    \label{fig:c0_ratio}
\end{figure}

This expression applies for $z\gg a$. Figure~\ref{fig:c0_ratio} compares the measured ratios with this strict NLO expansion. The measured ratios lie below their NLO counterparts, indicating a faster decrease of the zero-momentum matrix element between adjacent separations. These ratios directly probe the changes in the reference denominator when the transition distance is increased from \(2a\) to \(3a\) and from \(3a\) to \(4a\).

\begin{figure*}[!tp]
    \centering
    \includegraphics[width=0.47\linewidth]{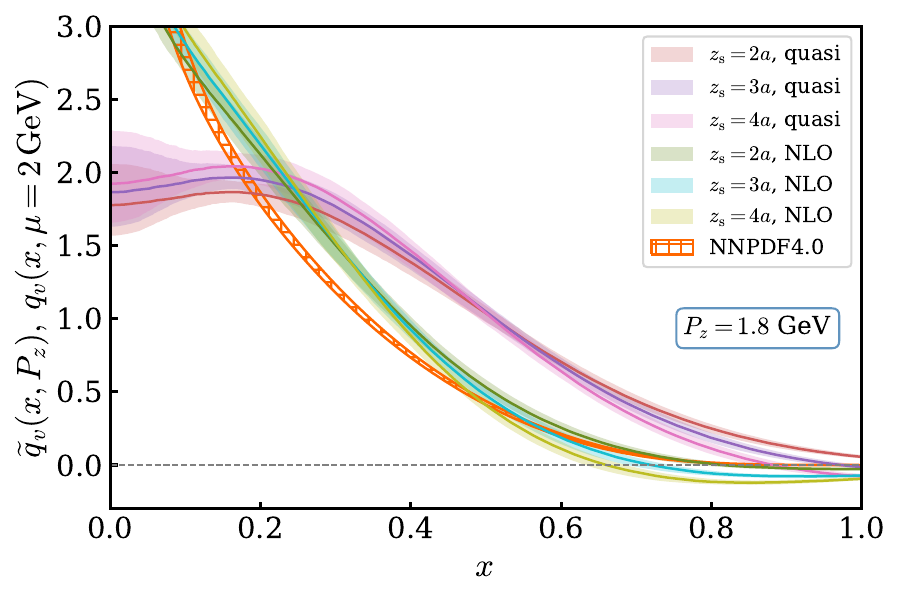}
    \includegraphics[width=0.47\linewidth]{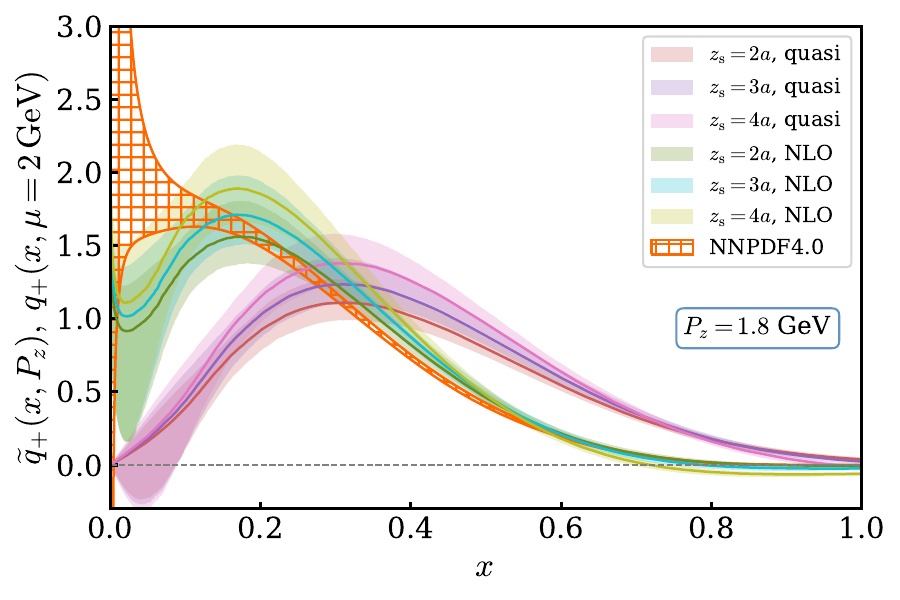} \\
    \includegraphics[width=0.47\linewidth]{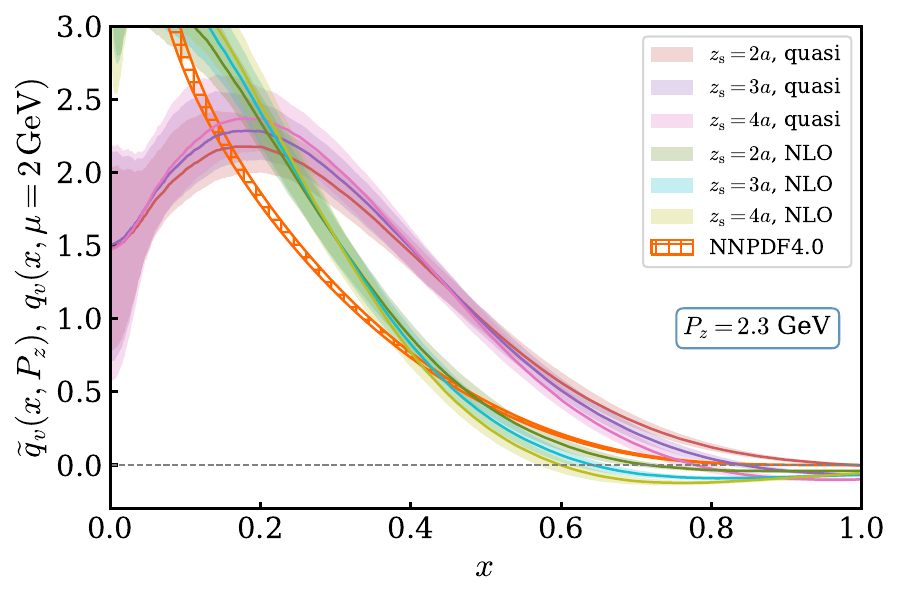}
    \includegraphics[width=0.47\linewidth]{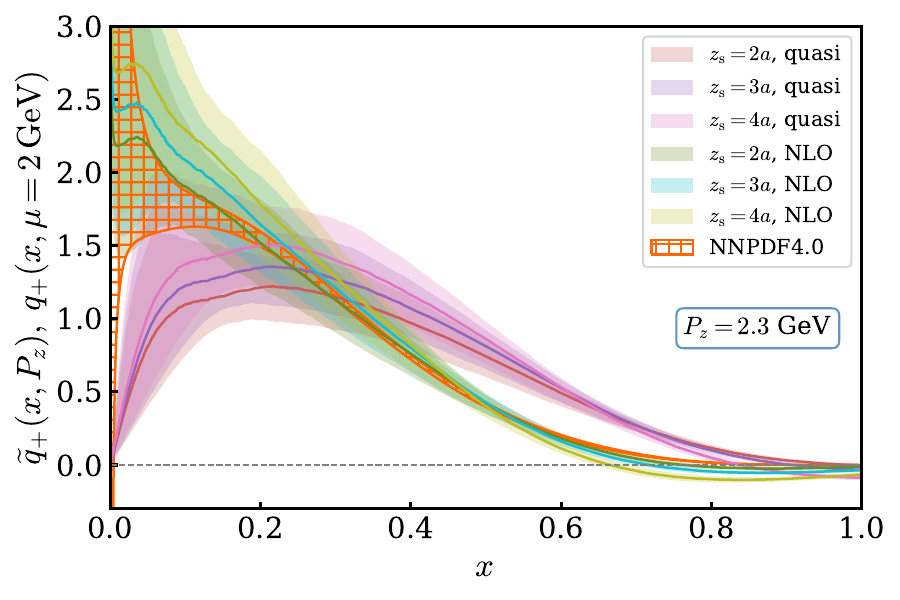}
    \caption{Transition-distance dependence of the isovector hybrid quasi-PDFs and NLO-matched light-cone PDFs. Top: $P_z=1.8$~GeV; bottom: $P_z=2.3$~GeV. Left: $q_v$; right: $q_+$. The three choices $\zs=2a$, $3a$, and $4a$ are compared before and after fixed-order NLO matching to $\overline{\rm MS}$ at $\mu=2$~GeV. Bands show the propagated uncertainties, and the orange cross-hatched band shows NNPDF4.0.}
    \label{fig:hybrid_zs_quasi_matched}
\end{figure*}

Although $C_0$ is not supplied as a separate numerical input to the PDF reconstruction, it describes the perturbative zero-momentum denominator used in deriving the hybrid matching coefficient. The $\zs$-dependent conversion term therefore encodes the short-distance separation dependence predicted by $C_0$. A mismatch between the measured ratios and their NLO counterparts can leave residual $\zs$ dependence after matching. Figure~\ref{fig:hybrid_zs_quasi_matched} compares the quasi-PDFs and fixed-order NLO light-cone PDFs for $\zs=2a$, $3a$, and $4a$ at $P_z=1.8$ and $P_z=2.3$~GeV. A visible transition-distance dependence is present in the quasi-PDFs and persists after matching. This does not contradict the short-distance agreement in Fig.~\ref{fig:hybrid_zs_coordinate}, since the quasi-PDF also depends on matrix elements at larger separations, which are affected by the mismatch between $C_0(z_s)$ and the zero-momentum matrix elements at $z_s$. The present diagnostic does not distinguish missing higher-order perturbative terms from $z^2\Lambda_{\rm QCD}^2$ corrections or finite-$a$ effects such as $a^2/z^2$. For the hybrid results presented in the main text, we use $\zs=2a=0.152$~fm, which is the closest to the value predicted by the NLO short-distance coefficient. Since $2a$ is only two lattice spacings, this choice does not establish a parametrically controlled hybrid window. These observations reveal residual transition-distance dependence in the current finite-order CG hybrid implementation, which needs further improvement, including higher-order perturbative corrections and finer lattice spacings.

\section{Perturbative matching kernels in Coulomb gauge}
\label{app:matching}

We collect the one-loop matching coefficients for the temporal Dirac structure $\Gamma=\gamma^t$ and the renormalization prescriptions used in this work. These coefficients enter the matching formula in Eq.~\eqref{eq:factorization}. We use $\xi=x/y$ and $p_z=|y|P_z$. For the hybrid prescription, the kernel also depends on the dimensionless combination $\zs p_z$.

The matching coefficient is expanded to NLO as
\begin{equation}
    C_{\gamma^t}^{\rm R}\!\left(\xi,\frac{\mu}{p_z}\right)
    =\delta(1-\xi)
    -\frac{\alphas(\mu)C_F}{2\pi}
     C_{\gamma^t, {\rm R}}^{(1)}\!\left(\xi,\frac{\mu}{p_z}\right)
    +\mathcal O(\alphas^2),
\end{equation}
with $C_F=4/3$. The common ratio-scheme coefficient entering both prescriptions reads~\cite{Gao:2023lny}
\begin{equation}
\begin{aligned}
    &C^{(1)}_{\rm ratio}\!\left(\xi,\frac{\mu}{p_z}\right)
     =\left[\frac{1+\xi^2}{1-\xi}\ln\frac{4p_z^2}{\mu^2}
       +\xi-1\right]_{+(1)}^{[0,1]}\\
    &\quad+\biggl\{\frac{1+\xi^2}{1-\xi}
       \bigl[\operatorname{sgn}(\xi)\ln|\xi|
       +\operatorname{sgn}(1-\xi)\ln|1-\xi|\bigr]\\
    &\quad+\operatorname{sgn}(\xi)
       +\frac{3\xi-1}{\xi-1}
        \frac{\tan^{-1}\!\left(\frac{\sqrt{1-2\xi}}{|\xi|}\right)}
             {\sqrt{1-2\xi}}-\frac{3}{2|1-\xi|}\biggr\}_{+(1)}^{(-\infty,\infty)}.
\end{aligned}
\end{equation}
For the $\overline{\rm MS}$-renormalized CG quasi-PDF, the one-loop coefficient is~\cite{Gao:2023lny, Gao:2026hix}
\begin{equation}
\begin{aligned}
    &C_{\gamma^t,\overline{\rm MS}}^{(1)}\!\left(\xi,\frac{\mu}{p_z}\right)
     =C_{\rm ratio}^{(1)}\!\left(\xi,\frac{\mu}{p_z}\right)
      +\frac{1}{2|1-\xi|}\\
    &\quad+\frac{1}{2}\delta(1-\xi)
      \left[1+\ln\!\left(\frac{4p_z^2}{\mu^2}\right)
      -\int_0^2\frac{{\rm d}\xi'}{|1-\xi'|}\right].
\end{aligned}
\label{eq:msbar_gamma_t_kernel}
\end{equation}
For the hybrid-renormalized CG quasi-PDF, the corresponding coefficient is
\begin{equation}
    C_{\gamma^t,{\rm hyb}}^{(1)}\!\left(\xi,\frac{\mu}{p_z},\zs p_z\right)
    =C_{\rm ratio}^{(1)}\!\left(\xi,\frac{\mu}{p_z}\right)
    +\delta C_{\rm hyb}^{(1)}(\xi,\zs p_z),
\end{equation}
where the hybrid-scheme correction is
\begin{equation}
    \delta C^{(1)}_{{\rm hyb}}(\xi, \zs p_z) = \left[ \frac{1}{2|1-\xi|} - \frac{\operatorname{Si}\left((1-\xi) \zs p_z\right)}{\pi (1-\xi)} \right]_{+(1)}^{(-\infty,\infty)}.
\end{equation}
Here, $\operatorname{Si}(x)=\int_0^x {\rm d}u \;(\sin u/u)$ is the sine integral function. The plus distribution prescription $\big[f(\xi)\big]_{+(1)}^{\cal D}$ is defined by
\begin{equation}
    \int_{\cal D} {\rm d}\xi\,\big[f(\xi)\big]_{+(1)}^{\cal D}\,g(\xi) = \int_{\cal D} {\rm d} \xi\,f(\xi)\,\big[g(\xi) - g(1)\big],
\end{equation}
which regulates the soft singularity at $\xi=1$. The superscript on $[\,\cdot\,]_{+(1)}^{\cal D}$ specifies the integration domain on which the plus prescription is applied.

\section{Implementation of renormalization-group resummation}
\label{app:rgr}

\subsection{Evolution and matching}

This appendix specifies the evolution used to obtain the NLO+NLL distributions discussed in Sec.~\ref{subsec:matching}. We denote the nonsinglet evolution operator from $\mu_0$ to $\mu$ by $U_\eta(\mu,\mu_0)$, where $\eta=v,+$ labels the valence and full-quark combinations. It satisfies
\begin{equation}
    \frac{\partial U_\eta(\mu,\mu_0)}{\partial\ln\mu^2}
    =P_\eta(\alphas(\mu))\otimes U_\eta(\mu,\mu_0),
\end{equation}
with the initial condition $U_\eta(\mu_0,\mu_0)=I$. The NLL evolution is generated by
\begin{equation}
    P_\eta(\alphas)
    =\frac{\alphas}{2\pi}P_{qq}^{(0)}
     +\left(\frac{\alphas}{2\pi}\right)^2P_\eta^{(1)}.
\end{equation}
Here $P_{qq}^{(0)}(z)=C_F[(1+z^2)/(1-z)]_+$, while $P_v^{(1)}=P_{\rm ns}^{(1)-}$ and $P_+^{(1)}=P_{\rm ns}^{(1)+}$. Their explicit expressions are given in Appendix~\ref{app:splitting_functions}. We use the two-loop running coupling and implement the evolution on the physical momentum-fraction interval through stepwise matrix exponentials. 

For each target $x>0$, we first evaluate the NLO matching relation in Eq.~\eqref{eq:factorization} at the common scale $\mu_0=2\kappa xP_z$, using the corresponding kernel specified in Appendix~\ref{app:matching}, to obtain $q_{\eta,{\rm NLO}}^{\overline{\rm MS}}(u,\mu_0)$. The resummed distribution is then
\begin{equation}
    q_{\eta,{\rm NLO+NLL}}^{\overline{\rm MS}}(x,\mu)
    =\int_x^1\frac{{\rm d}u}{u}\,
      U_\eta\!\left(\frac{x}{u};\mu,\mu_0\right)
      q_{\eta,{\rm NLO}}^{\overline{\rm MS}}(u,\mu_0).
    \label{eq:final_matching}
\end{equation}
Here $\mu_0=2\kappa xP_z$ is held fixed throughout the intermediate convolutions for each target $x$. The matching convolution in Eq.~\eqref{eq:factorization} extends over the full quasi-PDF support, whereas DGLAP evolution acts on the physical interval $0<u<1$.

\subsection{Two-loop nonsinglet splitting functions}
\label{app:splitting_functions}

The two-loop nonsinglet splitting functions entering the NLL evolution differ between the valence and full-quark channels~\cite{Curci:1980uw}. We denote them by $P_{\rm ns}^{(1)-}$ and $P_{\rm ns}^{(1)+}$, respectively, and write
\begin{equation}
    P_{\rm common}(z)=C_F\bigl[
      C_FP_F(z)+\tfrac12 C_AP_A(z)+n_fT_FP_{n_f}(z)
    \bigr],
\end{equation}
\begin{equation}
    P_{\rm ns}^{(1)\pm}(z)
    =P_{\rm common}(z)\pm\bigl(C_F^2-\tfrac12 C_F C_A\bigr)\bar P(z),
\end{equation}
where the color factors are $C_F=4/3$, $C_A=3$, $T_F=1/2$, and $n_f=3$ is the number of active quark flavors. The constituent functions are explicitly given by
\begin{equation}
\begin{aligned}
    P_F(z) &=-2\frac{1+z^2}{1-z}\ln z\ln(1-z)\\
    &\quad-\left(\frac{3}{1-z}+2z\right)\ln z
      -\frac{1+z}{2}\ln^2z-5(1-z),\\
    P_A(z) &=\frac{1+z^2}{1-z}\biggl(\ln^2z
      +\frac{11}{3}\ln z+\frac{67}{9}
      -\frac{\pi^2}{3}\biggr)\\
    &\quad+2(1+z)\ln z+\frac{40}{3}(1-z),\\
    P_{n_f}(z) &=\frac{2}{3}\biggl[
      \frac{1+z^2}{1-z}\left(-\ln z-\frac{5}{3}\right)
      -2(1-z)\biggr],\\
    \bar P(z) &=4(1-z)+2(1+z)\ln z\\
    &\quad+\frac{2(1+z^2)}{1+z}
      \biggl[\ln z\bigl(\ln z-2\ln(1+z)\bigr)\\
    &\qquad+\text{Li}_2\!\left(-\frac{1}{z}\right)
      -\text{Li}_2(-z)\biggr],
\end{aligned}
\end{equation}
where $\text{Li}_2(z)$ denotes the dilogarithm function.

\bibliographystyle{apsrev4-1}
\bibliography{ref.bib}

\end{document}